# Security in Wireless Sensor Networks

**Jaydip Sen**
Department of Computer Science & Engineering, National Institute of Science & Technology, INDIA
e-mail: Jaydip.Sen@acm.org

**Abstract**

Wireless sensor networks (WSNs) have attracted a lot of interest over the last decade in wireless and mobile computing research community. Applications of WSNs are numerous and growing, which range from indoor deployment scenarios in the home and office to outdoor deployment in adversary's territory in a tactical battleground. However, due to distributed nature and their deployment in remote areas, these networks are vulnerable to numerous security threats that can adversely affect their performance. This problem is more critical if the network is deployed for some mission-critical applications such as in a tactical battlefield. Random failure of nodes is also very likely in real-life deployment scenarios. Due to resource constraints in the sensor nodes, traditional security mechanisms with large overhead of computation and communication are infeasible in WSNs. Design and implementation of secure WSNs is, therefore, a particularly challenging task. This chapter provides a comprehensive discussion on the state of the art in security technologies for WSNs. It identifies various possible attacks at different layers of the communication protocol stack in a typical WSN and presents their possible countermeasures. A brief discussion on the future direction of research in WSN security is also included.

## 1. Introduction

**W**ireless sensor networks (WSNs) consist of hundreds or even thousands of small devices each with sensing, processing, and communication capabilities to monitor the real-world environment. They are envisioned to play an important role in a wide variety of areas ranging from critical military surveillance applications to forest fire monitoring and building security monitoring in the near future [1]. In these networks, a large number of sensor nodes are deployed to monitor a vast field, where the operational conditions are most often harsh or even hostile. However, the nodes in WSNs have severe resource constraints due to their lack of processing power, limited memory and energy. Since these networks are usually deployed in remote places and left unattended, they should be equipped with security mechanisms to defend against attacks such as node capture, physical tampering, eavesdropping, denial of service, etc. Unfortunately, traditional security mechanisms with high overhead are not feasible for resource constrained sensor nodes. The researchers in WSN security have proposed various security schemes which are optimized for these networks with resource constraints. A number of secure and efficient routing protocols [2 - 5], secure data aggregation protocols [6 - 11] etc. has been proposed by several researchers in WSN security.

In addition to traditional security issues like secure routing and secure data aggregation, security mechanisms deployed in WSNs also should involve collaborations among the nodes due to the decentralized nature of the networks and absence of any infrastructure. In real-world WSNs, the nodes can not be assumed to be trustworthy apriori. Researchers have therefore, focused on building a sensor trust model to solve the problems which are beyond the capabilities of traditional cryptographic mechanisms [12 - 19]. Since in most cases, the sensor nodes are unattended and physically insecure, vulnerability to physical attack is an important issue in WSNs. A number of propositions exist in the literature for defense against physical attack on sensor nodes [20 - 29].

In this chapter, we present a comprehensive overview of various security issues in WSNs. First we outline the constraints of WSNs, security requirements in these networks, and various possible attacks and the corresponding countermeasures. Then a holistic view of the security issues is presented. These issues are classified into six categories: cryptography, key management, secure routing, secure data aggregation, intrusion detection and trust management. The advantages and disadvantages of various security protocols are discussed, compared and evaluated. Some open research issues in each of these areas are also discussed.

The remainder of the chapter is organized as follows. In Section 2, various constraints in WSNs are discussed. Section 3 presents the security requirements in WSNs. Section 4 discusses various attacks that can be launched on WSNs. Section 5 presents the numerous countermeasures for all possible attacks on WSNs. Finally Section 6 concludes the paper highlighting some future directions of research in WSN security.

## 2. Constraints in Wireless Sensor Networks

A WSN consists of a large number of sensor nodes that are inherently resource-constrained devices. These nodes have limited processing capability, very low storage capacity, and constrained communication bandwidth. These constraints are due to limited energy and physical size of the sensor nodes. Due to these constraints, it is difficult to directly employ the conventional security mechanisms in WSNs. In order to optimize the conventional security algorithms for WSNs, it is necessary to be aware about the constraints of sensor nodes [30]. Some of the major constraints of a WSN are listed below.

*Energy constraints*: Energy is the biggest constraint for a WSN. In general, energy consumption in sensor nodes can be categorized in three parts: (i) energy for the sensor transducer, (ii) energy for communication among sensor nodes, and (iii) energy for microprocessor computation. The study in [31] found that each bit transmitted in WSNs consumes about as much power as executing 800 to 1000 instructions. Thus, communication is more costly than computation in WSNs. Any message expansion caused by security mechanisms comes at a significant cost. Further, higher security levels in WSNs usually correspond to more energy consumption for cryptographic functions. Thus, WSNs could be divided into different security levels depending on energy cost [32, 33].

*Memory limitations*: A sensor is a tiny device with only a small amount of memory and storage space. Memory is a sensor node usually includes flash memory and RAM. Flash memory is used for storing downloaded application code and RAM is used for storing application programs, sensor data, and intermediate results of computations. There is usually not enough space to run complicated algorithms after loading the OS and application code. In the SmartDust project, for example, TinyOS consumes about 4K bytes of instructions, leaving only 4500 bytes for running security algorithms and applications [31]. A common sensor type- TelosB- has a 16-bit, 8 MHz RISC CPU with only 10K RAM, 48K program memory, and 1024K flash storage [34]. The current security algorithms are therefore, infeasible in these sensors [35].

*Unreliable communication*: Unreliable communication is another serious threat to sensor security. Normally the packet-based routing of sensor networks is based on connectionless protocols and thus inherently unreliable. Packets may get damaged due to channel errors or may get dropped at highly congested nodes. Furthermore, the unreliable wireless communication channel may also lead to damaged or corrupted packets. Higher error rate also mandates robust error handling schemes to be implemented leading to higher overhead. In certain situation even if the channel is reliable, the communication may not be so. This is due to the broadcast nature of wireless communication, as the packets may collide in transit and may need retransmission [1].

*Higher latency in communication*: In a WSN, multi-hop routing, network congestion and processing in the intermediate nodes may lead to higher latency in packet transmission. This makes synchronization very difficult to achieve. The synchronization issues may sometimes be very critical in security as some security mechanisms may rely on critical event reports and cryptographic key distribution [36].

*Unattended operation of networks*: In most cases, the nodes in a WSN are deployed in remote regions and are left unattended. The likelihood that a sensor encounters a physical attack in such an environment is therefore, very high. Remote management of a WSN makes it virtually impossible to detect physical tampering. This makes security in WSNs a particularly difficult task.

## 3. Security Requirements in Wireless Sensor Networks

A WSN is a special type of network. It shares some commonalities with a typical computer network, but also exhibits many characteristics which are unique to it. The security services in a WSN should protect the information communicated over the network and the resources from attacks and misbehavior of nodes. The most important security requirements in WSN are listed below:

*Data confidentiality*: The security mechanism should ensure that no message in the network is understood by anyone except intended recipient. In a WSN, the issue of confidentiality should address the following requirements [30, 35]: (i) a sensor node should not allow its readings to be accessed by its neighbors unless they are authorized to do so, (ii) key distribution mechanism should be extremely robust, (iii) public information such as sensor identities, and public keys of the nodes should also be encrypted in certain cases to protect against traffic analysis attacks.

*Data integrity*: The mechanism should ensure that no message can be altered by an entity as it traverses from the sender to the recipient.

*Availability*: This requirements ensures that the services of a WSN should be available always even in presence of an internal or external attacks such as a denial of service attack (DoS). Different approaches have been proposed by researchers to achieve this goal. While some mechanisms make use of additional communication among nodes, others propose use of a central access control system to ensure successful delivery of every message to its recipient.

*Data freshness*: It implies that the data is recent and ensures that no adversary can replay old messages. This requirement is especially important when the WSN nodes use shared-keys for message communication, where a potential adversary can launch a replay attack using the old key as the new key is being refreshed and propagated to all the nodes in the WSN. A *nonce* or time-specific counter may be added to each packet to check the freshness of the packet.

*Self-organization*: Each node in a WSN should be self-organizing and self-healing. This feature of a WSN also poses a great challenge to security. The dynamic nature of a WSN makes it sometimes impossible to deploy any pre-installed shared key mechanism among the nodes and the base station [37]. A number of key pre-distribution schemes have been proposed in the context of symmetric encryption [37 - 40]. However, for application of public-key cryptographic techniques an efficient mechanism for key-distribution is very much essential. It is desirable that the nodes in a WSN self-organize among themselves not only for multi-hop routing but also to carryout key management and developing trust relations.

*Secure localization*: In many situations, it becomes necessary to accurately and automatically locate each sensor node in a WSN. For example, a WSN designed to locate faults requires accurate locations of sensor nodes to identify the faults. A potential adversary can easily manipulate and provide false location information by reporting false signal strength, replaying messages etc. if the location information is not secured properly. The authors in [41] have described a technique called verifiable multi-lateration (VM). In multi-lateration, the position of a device is accurately computed from a series of known reference points. The authors have used authenticated ranging and distance bounding to ensure accurate location of a node. Because of the use of distance bounding, an attacking node can only increase its claimed distance from a reference point. However, to ensure location consistency, the attacker would also have to prove that its distance from another reference point is shorter. As it is not possible for the attacker to prove this, it is possible to detect the attacker. In [42], the authors have described a scheme called *secure range-independent localization* (SeRLoC). The scheme is a decentralized range-independent localization scheme. It is assumed that the locators are trustworthy and cannot be compromised by any attacker. A sensor computes its location by listening to the beacon information sent by each locator which includes the locator's location information. The beacon messages are encrypted using a shared global symmetric key that is pre-distributed in the sensor nodes. Using the information from all the beacons that a sensor node receives, it computes its approximate location based on the coordinates of the locators. The sensor node then computes an overlapping antenna region using a majority vote scheme. The final location of the sensor node is determined by computing the center of gravity of the overlapping antenna region.

*Time synchronization*: Most of the applications in sensor networks require time synchronization. Any security mechanism for WSN should also be time-synchronized. A collaborative WSN may require synchronization among a group of sensors. In [43], a set of secure synchronization protocols have been proposed.

*Authentication*: It ensures that the communicating node is the one that it claims to be. An adversary can not only modify data packets but also can change a packet stream by injecting fabricated packets. It is, therefore, essential for a receiver to have a mechanism to verify that the received packets have indeed come from the actual sender node. In case of communication between two nodes, data authentication can be achieved through a *message authentication code* (MAC) computed from the shared secret key. A number of authentication schemes for WSNs have been proposed by researchers, most of which are for secure routing.

## 4. Security Vulnerabilities in Wireless Sensor Networks

WSNs are vulnerable to various types of attacks. These attacks can be broadly categorized as follows [44]:

- *Attacks on secrecy and authentication*: standard cryptographic techniques can protect the secrecy and authenticity of communication channels from outsider attacks such as eavesdropping, packet replay attacks, and modification or spoofing of packets.
- *Attacks on network availability*: attacks on availability are often referred to as *denial-of-service* (DoS) attacks. DoS attacks may target any layer of a sensor network.
- *Stealthy attack against service integrity*: in a stealthy attack, the goal of the attacker is to make the network accept a false data value. For example, an attacker compromises a sensor node and injects a false data value through that sensor node.

In these attacks, keeping the sensor network available for its intended use is essential. DoS attacks against WSNs may permit real-world damage to the health and safety of people [29]. The DoS attack usually refers to an adversary's attempt to disrupt, subvert, or destroy a network. However, a DoS attack can be any event that diminishes or eliminates a network's capacity to perform its expected functions [29].

### 4.1 Denial of service attacks

Wood et al. have defined a DoS attack as an event that diminishes or attempts to reduce a network's capacity to perform its expected function [29]. There are several standard techniques existing in the literature to cope with some of the more common denial of service attacks, although in a broader sense, development of a generic defense mechanism against DoS attacks is still an open problem. Moreover, most of the defense mechanisms require high computational overhead and hence not suitable for resource-constrained WSNs. Since DoS attacks in WSNs can sometimes prove very costly, researchers have spent a great deal of effort in identifying various types of such attacks, and devising strategies to defend against them. Some important types of DoS attacks in WSNs are discussed below.

### 4.1.1 Physical layer attacks

The physical layer is responsible for frequency selection, carrier frequency generation, signal detection, modulation, and data encryption [1]. As with any radio-based medium there exists the possibility of jamming in WSNs. There are two broad categories of attack on WSNs in the physical layer: (i) jamming and (ii) tampering. They are described as follows.

*Jamming*: it is a type of attack which interferes with the radio frequencies that the nodes use in a WSN for communication [29, 44]. A jamming source may be powerful enough to disrupt the entire network. Even with less powerful jamming sources, an adversary can potentially disrupt communication in the entire network by strategically distributing the jamming sources. An intermittent jamming may also prove detrimental [29].

*Tampering*: sensor networks typically operate in outdoor environments. Due to unattended and distributed nature, the nodes in a WSN are highly susceptible to physical attacks [45]. The physical attacks may cause irreversible damage to the nodes. The adversary can extract cryptographic keys from the captured node, tamper with its circuitry, modify the program codes or even replace it with a malicious sensor [28]. It has been shown that sensor nodes such as MICA2 motes can be compromised in less than one minute time [22].

### 4.1.2 Link layer attacks

The link layer is responsible for multiplexing of data-streams, data frame detection, medium access control, and error control [1]. Attacks at this layer include purposefully created collisions, resource exhaustion, and unfairness in allocation. A collision occurs when two nodes attempt to transmit on the same frequency simultaneously [29]. When packets collide, they are discarded and need to be retransmitted. An adversary may strategically cause collisions in specific packets such as ACK control messages. A possible result of such collisions is the costly exponential back-off. The adversary may simply violate the communication protocol and continuously transmit messages in an attempt to generate collisions. Repeated collisions can also be used by an attacker to cause resource exhaustion [29]. For example, a naïve link layer implementation may continuously attempt to retransmit the corrupted packets. Unless these retransmissions are detected early, the energy levels of the nodes would be exhausted quickly. Unfairness is a weak form of DoS attack [29]. An attacker may cause unfairness by intermittently using the above link layer attacks. In this case, the adversary causes degradation of real-time applications running on other nodes by intermittently disrupting their frame transmissions.

### 4.1.3 Network layer attacks

The network layer of WSNs is vulnerable to the different types of attacks such as: (i) spoofed routing information, (ii) selective packet forwarding, (iii) sinkhole, (iv) Sybil, (v) wormhole, (vi) blackhole and grayhole, (vii) HELLO flood, (viii) Byzantine, (ix) information disclosure, (x) acknowledgment spoofing etc. These attacks are described briefly in the following:

*Spoofed routing information*: the most direct attack against a routing protocol is to target the routing information in the network. An attacker may spoof, alter, or replay routing information to disrupt traffic in the network [46]. These disruptions include creation of routing loops, attracting or repelling network traffic from selected nodes, extending or shortening source routes, generating fake error messages, causing network partitioning, and increasing end-to-end latency.

*Selective forwarding*: in a multi-hop network like a WSN, for message communication all the nodes need to forward messages accurately. An attacker may compromise a node in such a way that it selectively forwards some messages and drops others [46].

*Sinkhole*: In a sinkhole attack, an attacker makes a compromised node look more attractive to its neighbors by forging the routing information [29, 46, 47]. The result is that the neighbor nodes choose the compromised node as the next-hop node to route their data through. This type of attack makes selective forwarding very simple as all traffic from a large area in the network would flow through the compromised node.

*Sybil attack*: it is an attack where one node presents more that one identity in a network. It was originally described as an attack intended to defeat the objective of redundancy mechanisms in distributed data storage systems in peer-to-peer networks [48]. Newsome et al describe this attack from the perspective of a WSN [47]. In addition to defeating distributed data storage systems, the Sybil attack is also effective against routing algorithms, data aggregation, voting, fair resource allocation, and foiling misbehavior detection. Regardless of the target (voting, routing, aggregation), the Sybil algorithm functions similarly. All of the techniques involve utilizing multiple identities. For instance, in a sensor network voting scheme, the Sybil attack might utilize multiple identities to generate additional "votes". Similarly, to attack the routing protocol, the Sybil attack would rely on a malicious node taking on the identity of multiple nodes, and thus routing multiple paths through a single malicious node.

*Wormhole*: a wormhole is low latency link between two portions of a network over which an attacker replays network messages [46]. This link may be established either by a single node forwarding messages between two adjacent but otherwise non-neighboring nodes or by a pair of nodes in different parts of the network communicating with each other. The latter case is closely related to sinkhole attack as an attacking node near the base station can provide a one-hop link to that base station via the other attacking node in a distant part of the network.

*Blackhole and Grayhole*: in the *blackhole attack*, a malicious node falsely advertises good paths (e.g., the shortest path or the most stable path) to the destination node during the path-finding process (in reactive routing protocols), or in the route update messages (in proactive routing protocols). The intention of the malicious node could be to hinder the path-finding process or to intercept all data packets being sent to the destination node concerned. A more delicate form of this attack is known as the *grayhole attack*, where the malicious node intermittently drops data packets thereby making its detection more difficult.

*HELLO flood*: most of the protocols that use *HELLO* packets make the naïve assumption that receiving such a packet implies that the sender is within the radio range of the receiver. An attacker may use a high-powered transmitter to fool a large number of nodes and make them believe that they are within its neighborhood [46]. Subsequently, the attacker node falsely broadcasts a shorter route to the base station, and all the nodes which received the *HELLO* packets, attempt to transmit to the attacker node. However, these nodes are out of the radio range of the attacker.

*Byzantine attack*: in this attack, a compromised node or a set of compromised nodes works in collusion and carries out attacks such as creating routing loops, forwarding packets in non-optimal routes, and selectively dropping packets [49]. Byzantine attacks are very difficult to detect , since under such attacks the networks usually do not exhibit any abnormal behavior.

*Information disclosure*: a compromised node may leak confidential or important information to unauthorized nodes in a network. Such information may include information regarding the network topology, geographic location of nodes, or optimal routes to authorized nodes in the network.

*Reource-depletion attack*: in this type of attack, a malicious node tries to deplete resources of other nodes in a network. The typical resources that are targeted are: battery power, bandwidth, and computational power. The attacks could be in the form of unnecessary requests for routes, very frequent generation of beacon packets, or forwarding of stale packets to other nodes.

*Acknowledgment spoofing*: some routing algorithms for WSNs require transmission of acknowledgment packets. An attacking node may overhear packet transmissions from its neighboring nodes and spoof the acknowledgments thereby providing false information to the nodes [46]. In this way, the attacker is able to disseminate wrong information in the network about the status of the nodes, since some acknowledgment may arrive from nodes which are not alive in reality.

In addition to above categories of attacks, there are various types of possible attacks on the routing protocols in WSNs. Most of the routing protocols in WSNs are vulnerable to attacks such as: routing table overflow, routing table poisoning, packet replication, route cache poisoning, rushing attacks etc. A comprehensive discussion on these attacks have been done in [50].

### 4.1.4 Transport layer attacks

The attacks that can be launched on the transport layer in a WSN are flooding attack and de-synchronization attack.

*Flooding*: Whenever a protocol is required to maintain state at either end of a connection, it becomes vulnerable to memory exhaustion through flooding [29]. An attacker may repeatedly make new connection request until the resources required by each connection are exhausted or reach a maximum limit. In either case, further legitimate requests will be ignored.

*De-synchronization*: De-synchronization refers to the disruption of an existing connection [29]. An attacker may, for example, repeatedly spoof messages to an end host causing the host to request the retransmission of missed frames. If timed correctly, an attacker may degrade or even prevent the ability of the end hosts to successfully exchange data causing them instead to waste energy attempting to recover from errors which never really exist. The possible DoS attacks and the corresponding countermeasures are listed in **Table 1**.

**Table 1**. Attacks on various layers of a WSN and their countermeasures

| Layer | Attacks | Defense |
| --- | --- | --- |
| Physical | Jamming | Spread-spectrum, priority messages, lower duty cycle, region mapping, mode change |
| Link | Collision | Error-correcting code |
|  | Exhaustion | Rate limitation |
|  | Unfairness | Small frames |
| Network | Spoofed routing information & Selective forwarding | Egress filtering, authentication, monitoring |
|  | Sinkhole | Redundancy probing |
|  | Sybil | Authentication, monitoring, redundancy |
|  | Wormhole | Authentication, probing |
|  | HELLO Flood | Authentication, packet leashes by using geographic and temporal info |
|  | Acknowledgment flooding | Authentication, verify the bi-directional link authentication |
| Transport | Flooding | Client puzzles |
|  | De-synchronization | Authentication |

Source: Y. Wang, G. Attebury, and B. Ramamurthy, IEEE Communications Surveys and Tutorials, Vol. 8, no. 2, pp. 2- 23, 2006.

### 4.2 Attacks on secrecy and authentication

There are different types of attacks under this category. They are described in Sections 4.2.1 through 4.2.2.

### 4.2.1 Node replication attack

In a *node replication attack*, an attacker attempts to add a node to an existing WSN by replication (i.e. copying) the node identifier of an already existing node in the network [51]. A node replicated and joined in the network in this manner can potentially cause severe disruption in message communication in the WSN by corrupting and forwarding the packets in wrong routes. This may also lead to network partitioning and communication of false sensor readings. In addition, if the attacker gains physical access to the entire network, it is possible for him to copy the cryptographic keys and use these keys for message communication from the replicated node. The attacker can also place the replicated node in strategic locations in the network so that he could easily manipulate a specific segment of the network, possibly causing a network partitioning.

### 4.2.2 Attacks on privacy

Since WSNs are capable of automatic data collection through efficient and strategic deployment of sensors, these networks are also vulnerable to potential abuse of these vast data sources. Privacy preservation of sensitive data in a WSN is particularly difficult challenge [52]. Moreover, an adversary may gather seemingly innocuous data to derive sensitive information if he knows how to aggregate data collected from multiple sensor nodes. This is in analogy to the *panda hunter problem*, where the hunter can accurately estimate the location of the panda by systematically monitoring the traffic [53].

The privacy preservation in WSNs is even more challenging since these networks make large volumes of information easily available through remote access mechanisms. Since the adversary need not be physically present to carryout the surveillance, the information gathering process can be done anonymously with a very low risk. In addition, remote access allows a single adversary to monitor multiple sites simultaneously [54].

Following are some of the common attacks on sensor data privacy [52, 54]:

- *Eavesdropping and passive monitoring*: This is most common and easiest form of attack on data privacy. If the messages are not protected by cryptographic mechanisms, the adversary could easily understand the contents. Packets containing control information in a WSN convey more information than accessible through the location server, Eavesdropping on these messages prove more effective for an adversary.
- *Traffic analysis*: In order to make an effective attack on privacy, eavesdropping should be combined with a traffic analysis. Through an effective analysis of traffic, an adversary can identify some sensor nodes with special roles and activities in a WSN. For example, a sudden increase in message communication between certain nodes signifies that those nodes have some specific activities and events to monitor. Deng et al have demonstrated two types of attacks that can identify the base station in a WSN without even underrating the contents of the packets being analyzed in traffic analysis [55].
- *Camouflage*: An adversary may compromise a sensor node in a WSN and later on use that node to masquerade a normal node in the network. This camouflaged node then may advertise false routing information and attract packets from other nodes for further forwarding. After the packets start arriving at the compromised node, it starts forwarding them to strategic nodes where privacy analysis on the packets may be carried out systematically.

It may be noted from the above discussion that WSNs are vulnerable to a number of attacks at all layers of the TCP/IP protocol stack. However, as pointed out by authors in [56], there may be other types of attacks possible which are not yet identified. Securing a WSN against all these attacks may be a quite challenging task.

## 5. Security Mechanisms for Wireless Sensor Networks

In this section, defense mechanism for combating various types of attacks on WSNs will be discussed. First, different cryptographic mechanisms for WSNs are presented. Both public key cryptography and symmetric key cryptographic techniques are discussed for WSN security. A number of key management protocols for WSNs are discussed next. Various methods of defending against DoS attacks, secure broadcasting mechanisms and various secure routing mechanisms are also discussed. In addition, various mechanisms for defending the Sybil attack, node replication attack, traffic analysis attacks, and attacks on sensor privacy are also presented. Finally, intrusion detection mechanisms for WSNs, secure data aggregation mechanisms and various trust management schemes for WSN security are discussed.

### 5.1 Cryptography in WSNs

Selecting the most appropriate cryptographic method is vital in WSNs as all security services are ensured by cryptography. Cryptographic methods used in WSNs should meet the constraints of sensor nodes and be evaluated by code size, data size, processing time, and power consumption. In this section, we focus on the selection of cryptography in WSNs. We discuss public key cryptography first, followed by symmetric key cryptography.

### 5.1.1 Public key cryptography in WSNs

Many researchers believe that the code size, data size, processing time, and power consumption make it undesirable for public key algorithm techniques, such as the *Diffie-Hellman key agreement protocol* [57] or RSA signatures [58], to be employed in WSNs.

**Table 2**. Public key cryptography: average ECC and RSA execution times

| Algorithm | Operation Time (s) |
|---|---|
| ECC secp160r1 | 0.81 |
| ECC secp224r1 | 2.19 |
| RSA-1024 public key e = $2^{16}$ + 1 | 0.43 |
| RSA-1024 private key (with Chinese Remainder Theorem) | 10.99 |
| RSA-2048 public key e = $2^{16}$ + 1 | 1.94 |
| RSA-2048 private key (with Chinese Remainder Theorem) | 83.26 |

Source: Y. Wang, G. Attebury, and B. Ramamurthy, IEEE Comm. Surveys and Tutorials, Vol. 8, no. 2, pp. 2-23, 2006

Public key algorithms such as RSA are computationally intensive and usually execute thousands or even millions of multiplication instructions to perform single cryptographic operation. Further, a microprocessor's public key algorithm efficiency is primarily determined by the number of clock cycles required to perform a multiplication instruction [30]. Brown et al. found that public key algorithms such as RSA usually require on the order of tens of seconds and up to minutes to perform encryption and decryption operations in resource-constrained wireless devices, which exposes a vulnerability to DoS attacks [59]. On the other hand, Carman et al found that it usually takes a microprocessor thousands of nano-joules to do a simple multiplication function with a 128-bit result [30]. In contrast, symmetric key cryptographic algorithms and hash functions consume much less computational energy than public key algorithms. For example, the encryption of a 1024-bit block consumes approximately 42mj on MC68328 DragonBall processor using RSA, and the estimated energy consumption for a 128-bit AES block is a much lower at 0.104 mj [30].

Studies have shown that it is feasible to apply public key cryptography to sensor networks by using the right selection of algorithms and associated parameters, optimization, and low power techniques [60 - 62]. The investigated public key algorithms include Rabin's Scheme [63], Ntru-Encrypt [64], RSA [58], and *elliptic curve cryptography* (ECC) [65, 66]. Most studies in the literature focus on RSA and ECC algorithms. The attraction of ECC is that it offers equal security for a far smaller key size, thereby reducing processing and communication overhead. For example, RSA with 1024-bit keys (RSA-1024) provides a currently accepted level of security for many applications and is equivalent in strength to ECC with 160-bit keys (ECC-160) [67]. To protect data beyond the year 2010, RSA Security recommends RSA-2048 as the new minimum key size, which is equivalent to ECC with 224-bit keys (ECC-224) [68]. **Table 2** summarizes the execution of ECC and RSA on an Atmel ATmega128 processor (used by Mica2 mote) [23]. The execution time is measured on average for a point multiplication in ECC and a modular exponential operation in RSA. ECC secp160r1 and secp224r1 are two standardized elliptic curves defined in [69]. As shown in **Table 2**, by using the small integer $e = 2^{16} + 1$ as the public key, RSA public key operation is slightly faster than ECC point multiplication. However, ECC point multiplication outperforms RSA private key operation by an order of magnitude. Since the RSA private key operations are too slow, they have limited use in sensor network applications. ECC has no such issues because both the public key operation and private key operation use the same point multiplication operations.

**Table 3**. Public Key Cryptography: Average Energy Costs of Digital Signature and Key Exchange in Millijoules (mJ)

| Algorithm | Signature | | Key Exchange | |
|---|---|---|---|---|
| | Sign | Verify | Client | Server |
| RSA-1024 | 304 | 11.9 | 15.4 | 304 |
| ECDSA-160 | 22.82 | 45.09 | 22.3 | 22.3 |
| RSA-2048 | 2302.7 | 53.7 | 57.2 | 2302.7 |
| ECDSA-224 | 61.54 | 121.98 | 60.4 | 60.4 |

Source: Y. Wang, G. Attebury, and B. Ramamurthy, IEEE Comm. Surveys and Tutorials, Vol 8, No. 2, pp. 2-23, 2006 [171]

Wander et al. investigated the energy cost of authentication and key exchange based on RSA and ECC cryptography on an Atmel ATmega128 processor [62]. The result is shown in **Table 3**. The ECC-based signature is generated and verified with the *elliptic curve digital signature algorithm* (ECDSA) [70]. The key exchange protocol is a simplified version of the SSL handshake, which involves two parties: a client initiating the communication and a server responding to the initiation [71]. The WSN is assumed to be administered by a central point with each sensor having a certificate signed by the central point's private key using an RSA or ECC signature. In the handshake process, the two parties verify each other's certificate and negotiate the session key to be used in the communication. As **Table 3** shows, compared with RSA cryptography at the same security level, ECDSA signatures are significantly cheaper than RSA signatures. Further, the ECC-based key exchange protocol outperforms the RSA-based key exchange protocol at the server side, and there is almost no difference in the energy cost for these two key exchange protocols at the client side. In addition, the relative performance advantage of ECC over RSA increases as the key size increases in terms of the execution time and energy cost. **Table 2** and **Table 3** indicate that ECC is more appropriate than RSA for use in sensor networks.

The implementation of RSA and ECC cryptography on Mica2 [31] nodes further proved that a public key-based protocol is viable for WSNs. In [72], Watro et al. have described a system named TinyPK where RSA system has been implemented on Mica2 motes using TinyOS development environment. The authors have demonstrated that authentication and key agreement protocol can be efficiently realized by this scheme in resource-constrained sensor nodes. Another scheme- TinyECC [73] based on ECC have been designed and implemented on Mica2. Similar work was also conducted by Malan et al. on ECC cryptography using a Mica2 mote [67]. In their work, ECC was used to distribute a single symmetric key for the link layer encryption provided by the TinySec module [74].

Although public key cryptography is possible in the sensor nodes, private key operations are still expensive. The assumptions mentioned in the literature [57-61] may not be satisfied in some applications. For example, the work in [57-61] concentrated on the public key operations only, assuming the private key operations will be performed by a base station or a third party. By selecting appropriate parameters, for example, using the small integer e = $2^{16}$ + 1 as the public key, the public key operation time can be extremely fast while the private key operation time does not change. The limitation of private key operation occurring only at a base station makes many security services using public key algorithms not available under these schemes. Such services include peer-to-peer authentication and secure data aggregation.

Table 4. Symmetric key cryptography: average RC5 and Skipjack execution times

| Algorithm | Operation Time (s) |
|---|---|
| Skipjack (C) [75] | 0.38 |
| RC5 (C, assembly) [76] | 0.26 |

Source: Y.Wang, G. Attebury, and B. Ramamurthy, IEEE Comm. Surveys and Tutorials, Vol. 8, No. 2, pp. 2-23, 2006

Table 5. Symmetric key cryptography: average energy for AES and SHA-1

| Algorithm | Operation Time (s) |
|---|---|
| SHA-1  (C) [77] | 5.9 µJ / byte |
| AES-128 Encryption / Decryption (assembly) [78] | 1.62 / 2.49 µJ / byte |

Source: Y. Wang, G. Attebury, and B. Ramamurthy, IEEE Comm. Surveys and Tutorials, Vol. 8, No. 2, pp. 2-23, 2006

In contrast, **Table 4** and **Table 5** show the execution time and energy cost of two symmetric cryptography protocols on an Atmet ATmega128 processor. In **Table 4**, the execution time was measured on a 64-bit block using an 80-bit key. From the **Table 4**, we can see that symmetric key cryptography is faster and consumes less energy when compared to public key cryptography.

**5.1.2 Symmetric key cryptography in WSNs**

Since most of the public key cryptographic mechanisms are computationally intensive, most of the research studies for WSNs focus on use of symmetric key cryptographic techniques. Symmetric key cryptographic mechanisms use a single shared key between the two communicating host which is used both for encryption and decryption. However, one major challenge for deployment of symmetric key cryptography is how to securely distribute the shared key between the two communicating hosts. This is a non-trivial problem since pre-distributing the key may not always be feasible.

Five popular encryption schemes RC4 [79], RC5 [76], IDEA [79], SHA-1 [77], and MD5 [79, 80], were evaluated on six different microprocessors ranging in word size from 8-bit (Atmel AVR) to 16-bit (Mitsubishi M16C) to 32-bit widths (StrongARM, XScale) in [81]. The execution time and code memory size were measured for each algorithm and platform. The experiments indicated uniform cryptographic cost for each encryption class and each architecture class. The impact of caches was negligible while Instruction Set Architecture (ISA) support is limited to specific effects on certain algorithms. Moreover, hashing algorithms (e.g., MD5, SHA-1) incur almost an order of magnitude higher overhead than encryption algorithms (e.g., RC4, RC5, and IDEA).

In [82], Law et al. evaluated two symmetric key algorithms: RC5 and TEA [83]. They further evaluated six block ciphers: RC5, RC6 [84], Rijndael [78], MISTY1 [85], KASUMI [86], and Camellia [87] on IAR Systems' MSP430F149 in [82]. The benchmark parameters were code, data memory, and CPU cycles. The evaluation results are presented in **Table 6**, in which the algorithms are ranked based on the key setup and encryption mode used. In both cases, the algorithms are optimized for speed of execution and memory space requirement and then ranked on the basis of their speed of execution, code size and data size in memory. The evaluation results showed that Rijndael is suitable for high security and energy efficiency requirements and MISTY1 is suitable for good storage and energy efficiency.

The performance of symmetric key cryptography is mainly decided by the following factors:

- *Embedded data bus width*: many encryption algorithms prefer 32-bit word arithmetic, but most embedded processors usually use an 8-bit or 16-bit wide data bus.
- *Instruction set*: the ISA has specific effects on certain algorithms. For example, most embedded processors do not support the variable-bit rotation instruction like *rotate bit left* (ROL) of the Intel architecture which greatly improves the performance of RC5.

**Table 6**. A summary of cipher performance on sensor nodes [81]

| Rank | By Key Steps | | | | | |
|---|---|---|---|---|---|---|
| | Size Optimized | | | Speed Optimized | | |
| | Code Memory | Data Memory | Speed | Code Memory | Data Memory | Speed |
| 1 | RC5-32 | MISTY1 | MISTY1 | RC6-32 | MISTY1 | MISTY1 |
| 2 | KASUMI | Rijndael | Rijndael | KASUMI | Rijndael | Rijndael |
| 3 | RC6-32 | KASUMI | KASUMI | RC5-32 | KASUMI | KASUMI |
| 4 | MISTY1 | RC6-32 | Camellia | MISTY1 | RC6-32 | Camellia |
| 5 | Rijndael | RC5-32 | RC5-32 | Rijndael | Camellia | RC5-32 |
| 6 | Camellia | Camellia | RC6-32 | Camellia | RC5-32 | RC6-32 |
| | By Encryption (CBC/CFB/OFB/CTR) | | | | | |
| | Size Optimized | | | Speed Optimized | | |
| Rank | Code Memory | Data Memory | Speed | Code Memory | Data Memory | Speed |
| 1 | RC5-32 | RC5-32 | Rijndael | RC6-32 | RC5-32 | Rijndael |
| 2 | RC6-32 | MISTY1 | MISTY1 | RC5-32 | MISTY1 | Camellia |
| 3 | MISTY1 | KASUMI | KASUMI | MISTY1 | KASUMI | MISTY1 |
| 4 | KASUMI | RC6-32 | Camellia | KASUMI | RC6-32 | RC5-32 |
| 5 | Rijndael | Rijndael | RC6-32 | Rijndael | Rijndael | KASUMI |
| 6 | Camellia | Camellia | RC5-32 | Camellia | Camellia | RC6-32 |

*Source: Ganesan, P., et al., In Proceedings of the 2$^{nd}$ ACM International Conference on Wireless Sensor Networks and Applications, ACM Press, New York, 151-59, 2003.*

Selecting the appropriate cryptography method for sensor nodes is fundamental to provide security services in WSNs. However, the decision depends on the computation and communication capability of the sensor nodes. Open research issues range from cryptographic algorithms to hardware design as described below:

- Recent studies on public key cryptography have demonstrated that public key operations may be practical in sensor networks. However, private key operations are still too expensive in terms of computation and energy cost to accomplish in a sensor node. The application of private key operations to sensor nodes needs to be studied further.
- Symmetric key cryptography is superior to public key cryptography in terms of speed and low energy cost. However, the key distribution schemes based on symmetric key cryptography are not perfect. Efficient and flexible key distribution schemes need to be designed.
- It is also likely that more powerful motes will need to be designed to support the increasing requirements on computation and communication in sensor nodes.

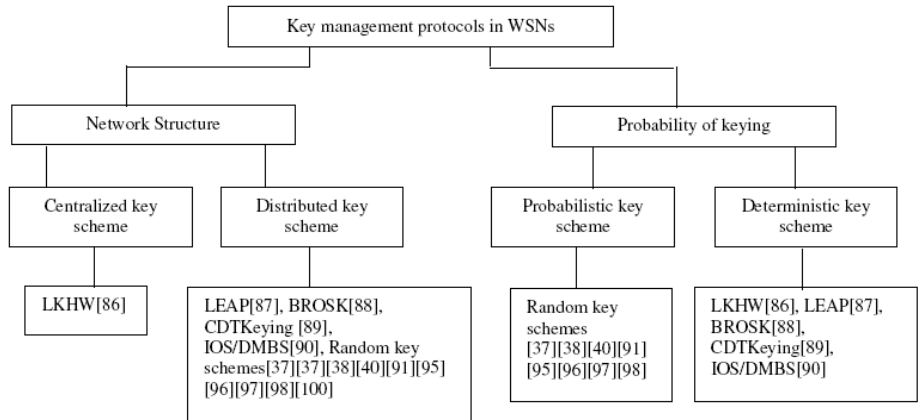

**Fig. 1.** Key management protococls in WSNs: a taxonomy [171]

**5.2 Key management protocols**

The area that has received maximum attention of the researchers in WSN security is key management. Key management is a core mechanism to ensure security in network services and applications in WSNs. The goal of key management is to establish the keys among the nodes in a secure and reliable manner. In addition, the key management scheme must support node addition and revocation in the network. Since the nodes in WSNs have computational and power constraints, the key management protocols for these networks must be extremely light-weight. Most of the existing key management protocols for WSNs are based on symmetric key cryptography because public key cryptographic techniques are in general computationally intensive. **Figure 1** presents a taxonomy of key management protocols in WSNs as described in [171]. In this Section, a brief overview of some of the most important key management protocols is given.

**5.2.1 Key management protocol based on network structure**

Depending on the underlying network structure, the key management protocols in WSNs may be centralized or distributed. In a centralized key management scheme, there is only one entity that controls the generation, re-generation, and distribution of keys. This entity is called *key distribution center* (KDC). The only protocol existing in the literature that is based on centralized key distribution is the LKHW scheme [88]. LKHW is based on *logical key hierarchy* (LKH). In this scheme, the base station is treated as a KDC and all keys are logically distributed in a tree rooted at the base station. The main drawback of this scheme is its single point of failure. If the central controller fails, the entire network and its security will be affected. The lack of scalability is another issue. Moreover, it does not provide data authentication. In the distributed key management protocols, different controllers are used to manage key-related activities. These protocols do not have the vulnerability of single point of failure and they allow better scalability. Most of the key management protocols existing in the literature are distributed in nature. These schemes fall either in deterministic or in probabilistic categories and are discussed in Section 5.2.2.1 and Section 5.2.2.2 respectively.

### 5.2.2 Key management protocols based on probability of key sharing

The key management protocols for WSNs may be classified on the probability of key sharing between a pair of sensor nodes. Depending of this probability the key management schemes may be either deterministic or probabilistic.

### 5.2.2.1 Deterministic key distribution schemes

The *localized encryption and authentication protocol* (LEAP) proposed by Zhu et al. [89] is a key management protocol for WSNs based on symmetric key algorithms. It uses different keying mechanisms for different packets depending on their security requirements. Four types of keys are established for each node: (i) an individual key shared with the base station (pre-distributed), (ii) a group of key shared by all the nodes in the network (pre-distributed), (iii) pair-wise key shared with immediate neighbor nodes, and (iv) a cluster key shared with multiple neighbor nodes. The pair-wise keys shared with immediate neighbor nodes are used to protect peer-to-peer communication and the cluster key is used for local broadcast.

It is assumed that the time required to attack a node is greater than the network establishment time, during which a node can detect all its intermediate neighbors. A common initial key is loaded into each node before deployment. Each node derives a master key which depends on the common key and its unique identifier. Nodes then exchange *HELLO* messages, which are authenticated by the receivers (since the common key and identifier are known, the master key of the neighbor can be computed). The nodes then compute a shared key based on their master keys. The common key is erased in all nodes after the completion of the key distribution process, and by assumption, no node has been compromised up to this point. Sine no adversary can get the common key, it is impossible to inject false data or decrypt the earlier exchange messages. Also, no node can later forge the master key of any other node. In this way, pair-wise shared keys are established between all immediate neighbors. The cluster key is established by a node after the pair-wise key establishment. A node generates a cluster key and sends it encrypted to each neighbor with its pair-wise shared key. The group key can be pre-loaded, but should be updated once any compromised node is detected. This could be done, in a naïve way, the base station's sending the new group key to each node using its individual key, or a hop-by-hop basis using cluster keys. Other sophisticated algorithms have been proposed for the same. Further, the authors [89] have proposed methods for establishing shared keys between multi-hop neighbors.

Lai et al. have proposed a *broadcast session key* (BROSK) negotiation protocol [90]. BROSK assumes a master key shared by all the nodes in the network. To establish a session key with its neighbor node $B$, a sensor node $A$ broadcasts a key negotiation message and both arrive at a shared session key. BROSK is a scalable and energy-efficient protocol.

Cametepe et al. have proposed a deterministic key distribution scheme for WSNs using combinatorial design theory [91]. The combinatorial design theory based pair-wise key pre-distribution (CDTKeying) scheme is based on block design techniques in combinatorics. It employs symmetric and generalized quadrangle design techniques. The scheme uses a finite projective plane of order $n$ (for prime power of $n$) to generate a symmetric design with parameters $n^2 + n + 1, n + 1, 1$. The design supports $n^2 + n + 1$ nodes and uses a key pool of size

$n^2 + n + 1$. It generates $n^2 + n + 1$ key chains of size $n + 1$ where every pair of key chains has exactly one key in common, and every key appears in exactly $n + 1$ key-chains. After the deployment, every pair of nodes finds exactly one common key. Thus, the probability of key sharing among a pair of sensor nodes is unity. The disadvantage of this proposition is that the parameter n has to be a prime power. Therefore, all network sizes can be supported for a fixed key chain size.

Lee et al. have proposed two combinatorial design theory based deterministic schemes: *ID-based one-way function scheme* (IOS) and *deterministic multiple space Bloms' scheme* (DMBS) [92]. They further discussed the use of combinatorial set systems in the design of deterministic key pre-distribution schemes for WSNs in [93].

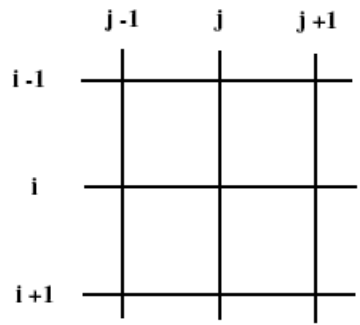

**Fig. 2.** The PIKE scheme: sensor nodes are organized in a two-dimensional space

Chan et al. have proposed a deterministic key management protocol to facilitate key establishment between every pair of neighboring nodes in a WSN [94]. In the mechanism, known as *peer intermediaries for key establishment* (PIKE), all $N$ sensor nodes are organized into a two-dimensional space as in **Figure 2**, where the coordinate of each node is $(x, y)$ for $x$, $y \in \{0, 1,.. \sqrt{N} – 1\}$. Each node shares unique pair-wise keys with $2(\sqrt{N} – 1)$ nodes that have the same $x$ or $y$ coordinate in the two-dimensional space. For two nodes with no common coordinate, an intermediate node, which has a common $x$ or $y$ coordinate with both nodes, is used as a router to forward a key from them. However, the communication overhead of the scheme is rather high because the secure connectivity is only $2 / \sqrt{N}$, which means that each node must establish a key for almost each of its neighbors through multi-link paths.

Huang et al. [95] have proposed a hybrid key establishment scheme that exploits the difference in computational and energy between a sensor node and the base station in a WSN. The authors argue that an individual sensor node possesses far less computational power and energy than a base station. In light of this, they propose placing the major cryptographic computations on the base station. On the sensor side, light-weight symmetric-key operations are deployed. Every sensor node and the base station mutually authenticate each other based on the ECC protocol. The proposed mechanism also uses certificates to establish the legitimacy of a public key. The certificates are based on an elliptic curve scheme. Such certificates are useful to verify the authenticity of sensor nodes.

Zhou and Fang [96] have developed a scalable key agreement protocol that uses a $t$-degree $(k + 1)$-variate symmetric polynomial to establish keys in a deterministic way.

#### 5.2.2.2 Probabilistic key distribution schemes

Most of the key management protocols for WSNs are probabilistic and distributed schemes. Eschenauer et al. have proposed a *random key pre-distribution* scheme for WSNs that relies on probabilistic key sharing among nodes of a *random graph* [37]. The mechanism has three phases: key pre-distribution, shared key discovery, and path key establishment. In the key pre-distribution phase, each sensor is equipped with a key ring stored in its memory. The key ring consists of *k* keys which are randomly drawn from a large pool of *P* keys. The association information of the key identifiers in the key ring and sensor identifier is also stored at the base station. Each sensor node shares a pair-wise key with the base station. In the shared key discovery phase, each sensor discovers its neighbors with which it shares keys. The authors have suggested two methods for this purpose. The simplest method is for each node to broadcast a list of identifiers of the keys in their key rings in plaintext allowing neighboring nodes to check whether they share a key. However, the adversary may observe the key-sharing patterns among sensors in this way. The second method uses the challenge-response technique to hide key-sharing patterns among nodes from an adversary. Finally, in the path key establishment phase, a path key is assigned for those sensor nodes within the communication range and not sharing a key, but connected by two or more links at the end of the second phase. If a node is compromised, the base station can send a message to all other sensors to revoke the compromised node's key ring. Re-keying follows the same procedure as revocation. The messages from the base station are signed by the pair-wise key shared by the base station and sensor nodes, thus ensuring that no adversary can forge a station. If a node is compromised, the attacker has a probability of approximately $k/P$ to attack any link successfully. Since $k << P$, it only affects a small number of sensor nodes.

Eschenauer et al.'s work can be considered as the basic random key management scheme. A number of additional key pre-distribution schemes have been proposed [38, 40, 97 - 100].

In the basic random key management scheme, any two neighbor nodes need to find a single common key from their key rings to establish a secure link in the key setup phase. However, Chan et al. observed that increasing the amount of key overlap in the key ring can increase the resilience of the network against node capture [38]. The authors have proposed *q-composite random key pre-distribution* scheme. It is required to share at least *q* common keys in the key setup phase to build a secure link between any two neighbor nodes. Further, they introduced a key update phase to enhance the basic random key management scheme. Suppose *A* has a secure link to *B* after the key setup phase and the secure key is *k* from the key pool *P*. Since *k* may be residing in the key ring memory of some other nodes in the network, the security of the link between *A* and *B* is jeopardized if any of those nodes are captured. Thus, it is better to update the communication key between *A* and *B* instead of using a key in the key pool. To address this problem, the authors have presented a multi-path key reinforcement for the key update. An adversary in this case has to eavesdrop on all the disjoint paths between node *A* and node *B* if he wants to reconstruct the communication key. The security of the scheme is further augmented by a random pair-wise key management scheme for node-to-node authentication.

To discover whether the key sets of two nodes have an intersection, usually both nodes need to broadcast their key indices or find common keys through a challenge-response procedure. Such methods have very high communication overhead. De Pietro et al. [98] improved the basic random key management scheme by associating the key indices of a node with its identity. For example, each node is assigned a pseudo-random number generator $g(x, y)$ and the key indices for the node are computed as $g(ID, i)$ for $i = 1, 2,….N$, where *ID* is the node identity. In this way, other nodes can find out which key is in its key set by checking its node identity.

Blundo et al. presented a *polynomial-based key pre-distribution* protocol for group key pre-distribution that can be adapted to WSNs [101]. The key setup server randomly generates a bivariate *t*-degree polynomial defined as:

$$f(x, y) = \sum_{i=0}^{t}\sum_{j=0}^{t} a_{ij} x^i y^j \tag{1}$$

The $t$-degree polynomial is defined over a finite field $\mathbf{F}_q$, where $q$ is a prime that is large enough to accommodate a cryptographic key. By choosing $a_{ij} = a_{ji}$, a symmetric polynomial is arrived at, i.e. $f(x, y) = f(y, x)$. Each sensor node is assumed to have a unique, integer-valued, non-zero identity. For each sensor node $u$, a polynomial share $f(u, y)$ is assigned, which means the coefficients of univariate polynomials $f(u, y)$ are loaded into the node $u$'s memory. When nodes $u$ and $v$ need to establish a shared key, they broadcast heir IDs. Subsequently, node $u$ can compute $f(u, v)$ by evaluating $f(u, y)$ at $y = v$, and node $v$ can also compute $f(v, u)$ by evaluating $f(v, y)$ at $y = u$. Due to the polynomial symmetry, the shared key between nodes $u$ and $v$ has been established as $K_{uv} = f(u, v) = f(v, u)$. A $t$-degree bivariate polynomial is also $(t + 1)$-secure. Therefore, an adversary must compromise no less than $(t + 1)$ nodes holding the shares of the same polynomial to reconstruct it.

Liu et al. have proposed a *polynomial pool-based key pre-distribution* (PPKP) scheme in [40]. The scheme also involves three phases: setup, direct key establishment, and path key establishment. In the setup phase, the setup server randomly generates a set $F$ of bivariate $t$-degree polynomials over the finite field $\mathbf{F}_q$. For each sensor node, the setup server picks up a subset of polynomials $F_i \subseteq F$ and assigns the polynomial shares of these polynomials to node $i$. In the direct key establishment stage [172], the sensor nodes finds a shared polynomial with other sensor nodes and then establish a pair-wise key using the polynomial-based key pre-distribution scheme discussed in [101]. The general framework based on polynomial pool-based pair-wise key pre-distribution can be applied in various ways. The authors [40] have provided two examples. In the random subset assignment strategy, during the setup phase, a setup server selects a random subset of polynomial shares to each sensor. In the second strategy-the grid-based key pre-distribution strategy- the setup server assigns a polynomial share to each node that is determined based on a grid structure. The grid-based key pre-distribution scheme is more resilient to a possible node compromise attack.

Du et al. have presented a *multiple-space key pre-distribution* (MSKP) scheme [97] which uses Blom's method [172]. The key difference between the schemes proposed in [40] and [97] is that the scheme in [40] is based on a set of bivariate $t$-degree polynomials, while the scheme in [97] is based on Blom's method. The proposed scheme allows any pair of nodes in a network to be able to find a pair-wise secret key. As long as no more than $\lambda$ nodes are compromised, the network is perfectly secure. To use Blom's method, during the pre-deployment phase, the base station first constructs a $(\lambda + 1) \times N$ matrix $G$ over a finite field $GF(q)$, where $N$ is the size of the network and $G$ is considered to be public information. Then the base station creates a random $(\lambda + 1) \times (\lambda + 1)$ symmetric matrix $D$ over $GF(q)$, and computes an $N \times (\lambda + 1)$ matrix $A = (D \cdot G)^T$, where $(D \cdot G)^T$ is the transpose of $D \cdot G$. Matrix $D$ needs to be kept secret, and should not be disclosed to adversaries. It is easy to verify that $A \cdot G$ is a symmetric matrix as follows.

$$A \cdot G = (D \cdot G)^T \cdot G = G^T \cdot D^T \cdot G = G^T \cdot D \cdot G = (A \cdot G)^T \tag{2}$$

Therefore, $K_{ij} = K_{ji}$. The idea is to use $K_{ij}$ (or $K_{ji}$) as the pair-wise key between node $i$ and node $j$. To carry out the above computation, in the pre-distribution phase for any sensor node k the following two steps are carried out: (i) the $k$-th row of matrix $A$ is stored at node $k$, and (ii) the $k$-th column of matrix $G$ is stored at node $k$. Then nodes $i$ and $j$ need to find the pair-wise key between them, they first exchange their columns of *G*, and then compute $K_{ij}$ and $K_{ji}$, respectively, using their private rows of *A*.

In the proposed scheme, each sensor node is loaded with *G* and *τ* distinct *D* matrices drawn from a large pool of *ω* symmetric matrices $D_1,\ldots D_\omega$ of size $(\lambda + 1) \times (\lambda + 1)$. For each $D_i$, calculate the matrix $A_i = (D_i \cdot G)^T$ and store the *j*-th row of $A_i$ at this node. After deployment, each node needs to discover whether it shares any space with neighbors. If they found out that they have a common space, the nodes can follow Blom's method to build a pair-wise key. The scheme is scalable and flexible. Moreover, it is substantially more resilient against node capture as compared to the scheme proposed in [40].

In the above scheme, each sensor node needs to keep many key materials such that a pair of nodes shares a key with a probability that can guarantee that the entire network is almost connected. This causes a large storage overhead on memory-constrained sensor nodes. Hwang et al. [39] proposed to enhance the basic random key management protocol [37] by reducing the amount of key-related materials required to be stored in each node, while guaranteeing a certain probability of sharing a key between two nodes. Their idea is to guarantee secure connectivity in the largest sub-component of the network rather than the entire network. The probability that two nodes have a key in common is reduced, but it is still large enough for the largest network component to be connected.

Hwang et al. extended the basic random key management scheme and proposed a cluster key grouping scheme [100]. They further analyzed the trade-offs involved between energy, memory, and security robustness.

In all the key management schemes discussed so far, the key materials are uniformly distributed in the entire terrain of a network. The uniform distribution makes the probability that two neighbor nodes share a direct key, called *secure connectivity*, rather small. Therefore, a lot of communication overhead is inevitable for the establishment of indirect keys. If some location information is known, two nearby sensor nodes can be preloaded with the same set of key materials. In this way, secure connectivity may be improved to a large extent.

In the *location-based key pre-distribution* (LBKP) scheme [102], the entire WSN is divided into many square cells. Each cell is associated with a unique *t*-degree bivariate polynomial. Each sensor node is pre-loaded with shares of the polynomials of its home cell and four other cells horizontally and vertically adjoining its home cells. After deployment, any two neighbor nodes can establish a pair-wise key if they have shares of the same polynomial. For example, in **Figure 3**, polynomial of cell $C_{33}$ is also assigned to cells $C_{32}$, $C_{34}$, $C_{23}$, and $C_{43}$. The polynomials of other cells are assigned in the same way. As a result, a node in $C_{33}$ has some polynomial information in common with other nodes in the shaded areas.

Du et al. [99] have also proposed a key pre-distribution scheme that uses network deployment knowledge. In the proposed scheme, the entire network is divided into many square cells. Each cell is assigned a subset key pool $S_{ij}$, $i = 1,\ldots u$ and $j = 1,\ldots v$ out of a global key pool *S*. Those subset key pools are set up such that the key pools of two neighbor cells will share a portion of keys. In each cell, the basic random key management scheme [37] is applied. Using the deployment knowledge- the information about the manner in which the nodes are deployed in the network- the scheme ensures that the value of the probability that a pair o neighboring nodes share a secret key is very high. The high value of the probability signifies that any pair of nodes in the network can establish secure communication sessions between them. The intelligent use of the deployment knowledge also ensures that the size of the key ring (i.e., the set of keys) held by a given node in the proposed scheme [99] is much smaller than that in the basic key management scheme proposed by Eschenauer and Gligor [37]. Hence, the scheme is very memory-efficient.

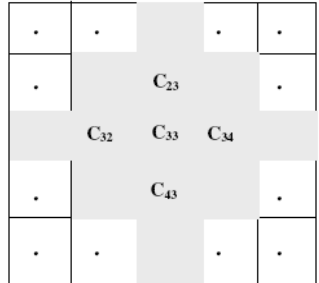

**Fig. 3.** The location-based key distribution (LBKP) scheme

Some of the above-mentioned key management schemes for WSNs are classified and compared in **Table 7**. Although a number of key management protocols have been proposed for WSNs, the design of key management protocols is still largely open to research. Some of the open research issues are discussed below.

**Table 7**. Classification and comparison of key management protocols in WSNs [103]

| Prot. Type | Protocol Name | Ref | Master Key | Pairwise Key | Path Key | Cluster Key | Scalabi-lity | Robust-ness | Proc. Load | Comm. Load | Storage Load |
|---|---|---|---|---|---|---|---|---|---|---|---|
| Deterministic | All pairwise | | NA | Yes | No | No | Low | Low | Low | Low | High |
| | LEAP | [89] | Yes | Yes | Yes | Yes | Good | Low | Low | Low | Low |
| | BROSK | [90] | Yes | Yes | No | No | Good | Low | Low | Low | Low |
| | LKHW | [88] | Yes | Yes | No | Yes | Fair | Low | Low | Low | Low |
| | CDTKeying | [91] | NA | Yes | No | No | Good | Good | Med | Med | High |
| | IOS & DMBS | [92] | NA | Yes | No | No | Good | Good | Med | Med | High |
| Probabilistic | Basic | [37] | NA | Yes | Yes | No | Good | Good | Med | Med | High |
| | q-composite | [38] | NA | Yes | No | No | Good | Good | Med | Med | High |
| | Polynomial based | [40] | NA | Yes | No | No | Good | Good | Med | Med | High |
| | Blom based | [97] | NA | Yes | No | No | Good | Good | Med | Med | High |
| | Deployment knowledge based | [99] | NA | Yes | No | No | Good | Good | Med | Med | High |
| | Cluster key grouping | [100] | NA | Yes | No | No | Good | Good | Med | Med | High |
| | Location based | [102] | NA | Yes | No | No | Good | Good | Med | Med | Med |

*Memory*: High security and lower overhead are two objectives that a key management protocol needs to achieve. Although there have been several proposals for key establishment in sensor networks, they can hardly address these two requirements. Strong security protocols usually require large amounts of memory cost, as well as high-speed processors and large power consumption. However, they cannot be easily supported due to the constraints on hardware resources of the sensor platform. It is well known that in wireless environment, transmission of one bit can consume more energy than computing one bit. In key management protocols, direct key establishment does not require communication or only a few rounds of one-hop communication, but indirect key establishment is performed over multi-hop communication. To reduce the multi-hop communication overhead, the probability of a direct key establishment between a pair of nodes should be as high as possible so that a secure connectivity among the nodes can be guaranteed. However, highly secure connectivity requires more key materials in each node, which is usually impractical, especially when the network size is large. Considering the above two issues, memory cost can be a major bottleneck in designing key management protocols in a WSN. How to reduce memory cost while still maintaining a certain level of security is a very important issue.

*End-to-end security*: The major merit of symmetric key cryptography is its computational efficiency. However, most current symmetric key schemes for WSNs aim at the link layer security- not the transport layer security- because it is impractical for each node to store a transport layer key for each of the other nodes in a network due to huge number of nodes. However, end-to-end communication at the transport layer is very common in many WSN applications. For example, to reduce unnecessary traffic, a fusion node can aggregate reports from many source nodes and forward a final report to the sink node. During this procedure, the reports between source nodes and the fusion node and the one between the fusion node and the sink node should be secured. In hostile environments, however, any node can be compromised. Of one of the intermediate nodes along a route is compromised, the message delivered along the route can be exposed or modified by the compromised node. Employing end-to-end security can effectively prevent message tampering by any malicious intermediate node. Compared with symmetric key technology, public key cryptography is expensive but has flexible manageability and supports end-to-end security. A more promising approach to key establishment in WSNs is to combine the merits of both symmetric key and public key techniques, in that each node is equipped with a public key system and relies on it to establish end-to-end symmetric keys with other nodes. To achieve this goal, a critical issue is to develop more efficient public key algorithms and their implementations so that they can be widely used on sensor platforms. How to prove the authenticity of public keys is another important problem. A malicious node can otherwise, impersonate any normal node by claiming its public key. Identity-based cryptography is a shortcut to avoid the problem. Currently, most identity-based cryptographic algorithms operate on elliptic curve-fields, and pairing over elliptic curves is widely used in the establishment of identity-based symmetric keys. However, the pairing operation is very costly, comparable to or even more expensive than RSA. Therefore, fast algorithms and implementations are the major tasks for the researchers.

*Efficient symmetric key algorithms*: There is still a demand for the development of more efficient symmetric key algorithms because encryption and authentication based on symmetric keys are very frequent in the security operations of sensor nodes. For example, in the link layer security protocol TinySec [74], each packet must be authenticated, and encryption can also be triggered if critical packets are transmitted. Therefore, fast and cost-efficient symmetric key algorithms should be developed.

*Key update and revocation*: Once a key has been established between two nodes, the key can act as a master key and be used to derive different sub-keys for many purposes (e.g., encryption and authentication0. If each key is used for a long time, it may be exposed due to cryptanalysis over the ciphertexts intercepted by adversaries. To protect the master key and those sub-keys from cryptanalysis, it is wise to update keys periodically. The period of update, however, is difficult to choose. Because the cryptanalysis capability of adversaries is unknown, it is very difficult to estimate how long it takes for adversaries to expose a key by cryptanalysis. If the key update period is too long, the corresponding key may also be exposed. If it is too short, frequent updates can incur large overhead. A related problem is key revocation. If one node is detected to be malicious, its key must be revoked. However, key revocation has not been thoroughly investigated. Although Chan et al. [104] proposed a distributed revocation protocol, it is only based on the random pair-wise key scheme [38], and cannot easily be generalized to fit other key establishment protocols.

*Node compromise*: Node compromise is the most detrimental attack on sensor networks. Because compromised nodes have all the authentic key materials, hey can result in very severe damage to WSN applications and cannot be detected easily. How to counteract node compromise remains an open problem. Most of the current security protocols attempt to minimize the adverse impact on the network due to a possible node compromise through careful protocol design such that the impact of node compromise can be restricted to a small area. However, a hardware approach is more promising. With advances in hardware design and manufacturing techniques, much stronger, tamper-resistant, and cheaper devices can be installed on the sensor platform to counteract node compromise.

**5.3 Defense against DoS attacks**

Various types of DoS attacks in WSNs have been discussed in Section 4. In this section, defense mechanisms for each of those attacks are presented in detail.

**5.3.1 Defense mechanisms in the physical layer**

Jamming attack may be defended by employing variations of spread-spectrum communication such as frequency hopping and code spreading [29]. *Frequency-hopping spread spectrum* (FHSS) is a method of transmitting signals by rapidly switching a carrier among many frequency channels using a pseudo-random sequence known to both the transmitter and the receiver. As a potential attacker would not be able to predict the frequency selection sequence, it will be impossible for him to jam the frequency being used at a given point of time. Code spreading is another technique for defending against jamming. However, it requires greater design complexity and energy and thus not very suitable for WSNs. In general, sensor devices are limited to single-frequency use and are highly susceptible to jamming attacks. One approach for tolerance against jamming attack in a WSN is to identify the jammed part of the network and effectively avoid it by routing around. Wood et al. [29] have proposed an approach where the nodes along the perimeter of a jammed region report their status to the neighbors and collectively the affected region is identified and packets are routed around it.

**5.3.2 Defense mechanisms in the link layer**

A typical defense against collision attack is the use of error-correcting codes [29]. Most codes work best with low levels of collisions such as those caused by environmental or probabilistic errors. However, these codes also add additional processing and communication overhead. It is reasonable to assume that an attacker will always be able to corrupt more than what can be corrected. Although it is possible to detect these malicious collisions, no complete defense mechanism against them is kwon today.

A possible solution for energy exhaustion attack is to apply a rate limiting MAC admission control. This would allow the network to ignore those requests that intend to exhaust the energy reserves of a node. A second technique is to use time-division multiplexing where each node is allotted a time slot in which it can transmit [29]. This eliminates the need of arbitration for each frame and can solve the indefinite postponement problem in a back-off algorithm. However, it is still susceptible to collisions.

The effect of unfairness caused by an attacker who intermittently launches link layer attacks can be lessened by use of small frames since it reduces the amount of time an attacker gets at his disposal to capture the communication channel [29]. However, this technique often reduces efficiency and is susceptible to further unfairness such as an attacker trying to retransmit quickly instead of randomly delaying.

### 5.3.3 Defense mechanisms in the network layer

A countermeasure against spoofing and alteration is to append a *message authentication code* (MAC) after the message. By adding a MAC to the message, the receivers can verify whether the messages have been spoofed or altered. To defend against replayed information, counters or time-stamps may be introduced in the messages [35]. A possible defense against selective forwarding attack is using multiple paths to send data [46]. A second defense is to detect the malicious node or assume it has failed and seek an alternative route.

Sen et al. have presented a cooperative detection scheme for identifying malicious packet dropping nodes in an ad hoc network [105]. The scheme exploits the redundancy in routing information in an ad hoc network to build a robust detection framework so that it works even in presence of transient network partitioning and Byzantine failure of nodes.

Hu et al. have proposed a novel and generic mechanism called *packet leashes* for detecting and defending against wormhole attacks [106]. In a wormhole attack, a malicious node eavesdrops on a series of packets, then tunnels them through a path in the network, and replays them. This is done in order to make a false representation of the distance between the two colluding nodes. It is also used, more generally, to disrupt the routing protocol by misleading the neighbor discovery process [46]. Hu et al. have presented a mechanism that employs directional antenna to combat wormhole attack [23]. Wang et al. have used a visualization approach to detect wormholes in a WSN [107]. In the mechanism proposed by the authors, a distance estimation is made between all the sensor nodes in a neighborhood. Using multi-dimensional scaling, a virtual layout of the network is then computed, and a surface smoothing strategy is used to adjust the round-off errors. Finally, the shape of the resulting virtual network is analyzed. If any wormhole exists, the shape of the network will bend and curve towards the wormhole; otherwise, the network will appear flat. Sen et al. have presented a security mechanism that can detect cooperative grayhole attacks in a wireless ad hoc and sensor network [108]. In this scheme, every node monitors the packet forwarding behavior of each of its neighbors and a global detection algorithm is employed to detect any routing misbehavior.

To defend against flooding DoS attack at the transport layer, Aura et al. have proposed using *client puzzles* [109], where each client should demonstrate its commitment to the connection by solving a puzzle. As an attacker does not have infinite resource, it will be impossible for him to create new connections fast enough to cause resource starvation on the serving node. A possible defense against de-synchronization attack is to enforce a mandatory requirement of authentication of all packets communicated between nodes [29]. If the authentication mechanism is secure, an attacker will be unable to send any spoofed messages.

Some mechanisms for secure multicasting and broadcasting in WSNs are now discussed.

### 5.3.3.1 Secure broadcasting and multicasting protocols

Multicasting and broadcasting techniques are used primarily to reduce the communication and management overhead of sending a single message to multiple receivers. In order to ensure that only legitimate group members receive the multicast and broadcast communication, appropriate authentication and encryption mechanisms must be in place. To handle this problem, several key management schemes have been devised: centralized group key management protocols, decentralized key management protocols, and distributed key management protocols [110]. First, we will discuss some generic security mechanisms for multicast and broadcast communication in wireless networks. Then we will present some of the well-known propositions specific to WSNs.

In the case of the centralized group key management protocols, a central authority is used to maintain the group. Decentralized management protocols, however, divide the task of group management amongst multiple nodes. In distributed key management protocols, the key management activity is distributed among a set of nodes rather than on a single node. In some cases, the entire group of nodes is responsible for key management [110].

An efficient way to distribute keys in a network is to use a logical key tree. Such techniques essentially fall under the category of centralized key management protocols. Some schemes have been developed for WSNs based on *logical key tree technique* [88,111, 112]. While centralized solutions are not always the most efficient ones, these mechanisms may sometimes be very effective for WSNs, as relatively heavier computations can be usually carried out in powerful base stations

Di Pietro et al. have proposed a directed diffusion-based multicast mechanism for WSNs that utilizes a logical key hierarchy [88]. In the logical hierarchy, a central key distributor is at the root of a tree, and the nodes in the network are the leaf level. The internal nodes of tree contain keys that are used in the re-keying process. The directed diffusion is an energy-efficient data dissemination technique for WSNs [113]. In directed diffusion, a query is transformed into an interest and then diffused throughout the network. The source node then starts collecting data from the network based on the propagated interest. The dissemination technique also sets up certain gradients designed to draw events toward the interest. The collected data is then sent back to the source along the reverse path of the interest propagation. The directed diffusion-based logical key hierarchy scheme as proposed by Di Pietro et al. allows nodes to join and leave groups. The key hierarchy is used to effectively re-establish keys for the nodes below the node that has left the group. When a node declares its intension to join a group, a key set is generated for the new node based on the keys within the existing key hierarchy.

Kaya et al. discuss the problem of multicast group management in [114], where the nodes in a network are grouped based on their locality and a security tree is constructed on the groups.

Lazos et al. have presented a tree-based key distribution scheme that is similar to the directed diffusion-based logical key hierarchy proposed by Di Pietro et al. [112]. In their proposed scheme, a routing-aware tree is constructed in which the leaf nodes are assigned keys based on all relay nodes above them. As the scheme takes advantage of routing information for construction the key hierarchy, it is more energy-efficient than routing schemes that arbitrarily arrange nodes into a routing tree.

In [111], the authors have proposed a mechanism that uses geographic location information for construction of a logical key hierarchy for secure multicast communication. The nodes, based on the geographical location information, are grouped into different clusters. The nodes within a cluster are able to reach each other with a single hop communication. Using the cluster information, a key hierarchy is constructed in a manner similar to that proposed in [112].

**5.4 Defense against attacks on routing protocols**

Many routing protocols have been proposed for WSNs. These protocols can be divided into three broad categories according to the network structure: (i) Flat structure-based routing, (ii) hierarchical structure-based routing, and (iii) location-based routing [115]. In flat-based routing, all nodes are typically assigned equal roles or functionality. In hierarchical-based routing, nodes play different roles in the network. In location-based routing, sensor node positions are used to route data in the network. One common location-based routing protocol is the *greedy perimeter stateless routing* (GPSR) [3]. It allows nodes to send packets to a region rather than a particular node. All these routing protocols are vulnerable to various types of attacks such as selective forwarding, sinkhole attack etc as mentioned in Section 4. Elaborate discussions on various types of attacks on the routing protocols in WSNs are given in [46] and [50]. A comparative analysis of some of the well known existing secure routing protocols for WSNs has been presented in [50].

The goal of a secure routing protocol for a WSN is to ensure the integrity, authentication, and availability of messages. Most of the existing secure routing algorithms for WSNs are based on symmetric key cryptography except the one described by Du et al. [116], which is based on public key cryptography. In this section, a number of security mechanisms for routing in WSNs are discussed in detail.

*µTESLA* (the "micro" version of the Timed, Efficient, Streaming, Loss-tolerant Authentication protocol) [9] and its extensions [117, 118] have been proposed to provide broadcast authentication for sensor networks. *µTESLA* is broadcast authentication protocol which was proposed by Perrig et al. for the *security protocols for sensor networks* (SPINS) protocol [35]. *µTESLA* introduces asymmetry through a delayed disclosure of symmetric keys resulting in an efficient broadcast authentication scheme. For its operation, it requires the base station and the sensor nodes to be loosely synchronized. In addition, each node must know an upper bound on the maximum synchronization error.

To send an authenticated packet, the base station simply computes a MAC on the packet with a key that is secret at that point of time. When a node gets a packet, it can verify that the corresponding MAC key was not yet disclosed by the base station. Because a receiving node is assured that the MAC key is known only to the base station, the receiving node is assured that no adversary could have altered the packet in transit. The node stores the packet in a buffer. At the time of key disclosure, the base station broadcasts the verification key to all its receivers. When a node receives the disclosed key, it can easily verify the correctness of the key. If the key is correct, the node can now use it to authenticate the packet stored in its buffer. Each MAC is a key from the key chain, generated by a public one-way function $F$. To generate the one-way key chain, the sender chooses the last key $K_n$ from the chain, and repeatedly applies $F$ to compute all other keys: $K_i = F(K_{i+1})$.

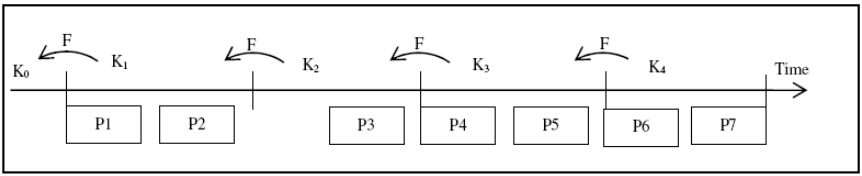

**Fig. 4.** Illustration of time-released key chain for source authentication [171]

**Figure 4** shows an example of *µTESLA*. The receiver node is loosely time synchronized and knows $K_0$ in an authenticated way. Packets $P_1$ and $P_2$ sent in interval 1 contain a MAC with a key $K_1$. Packet $P_3$ has a MAC using key $K_2$. If $P_4$, $P_5$, and $P_6$ are all lost, as well as the packet that disclosed the key $K_1$, the receiver cannot authenticate $P_1$, $P_2$, and $P_3$. In interval 4, the base station broadcasts the key $K_2$, which the nodes authenticate by verifying $K_0 = F(F(K_2))$, and hence know also $K_1 = F(K_2)$, so they can authenticate packets $P_1$, $P_2$ with $K_1$, and $P_3$ with $K_2$. SPINS limits the broadcasting capability to only the base station. If a node wants to broadcast authenticated data, the node has to broadcast the data through the base station. The data is first sent to the base station in an authenticated way. It is then broadcasted by the base station.

To bootstrap a new receiver, *µTESLA* depends on a *point-to-point authentication* mechanism in which a receiver sends a request message to the base station and the base station replies with a message containing all the necessary parameters. It may be noted that *µTESLA* requires the base station to unicast initial parameters to individual sensor nodes, and thus incurs a long delay to boot up a large-scale sensor network. Liu et al. propose a multi-level key chain scheme for broadcast authentication to overcome this deficiency [117, 118].

The basic idea in [117, 118] is to predetermine and broadcast the initial parameters required by *µTESLA* instead of using unicast-based message transmission. The simplest way is to pre-distribute the *µTESLA* parameters with a master key during the initialization of the sensor nodes. As a result, all sensor nodes have the key chain commitments and other necessary parameters once they are initialized, and are ready to use *µTESLA* as long as the staring time has passed. Furthermore, the authors have introduced a multi-level key chain scheme, in which the higher key chains are used to authenticate the commitments of the lower-level ones. However, the multi-level key chain suffers from possible DoS attacks during commitment distribution stage. Further, none of the *µTESLA* or multi-level key chain schemes is scalable in terms of the number of senders. In [119], a practical broadcast authentication protocol has been proposed to support a potentially large number of broadcast senders using *µTESLA* as a building block.

*µTESLA* provides broadcast authentication for base stations, but is not suitable for local broadcast authentication. This is because *µTESLA* does not provide immediate authentication. For every received packet, a node has to wait for one *µTESLA* interval to receive the MAC key used in computing the MAC for the packet. As a result, if *µTESLA* is used for local broadcast authentication, a message traversing l hops will take at least l *µTESLA* intervals to arrive at the destination. In addition, a sensor node has to buffer all unverified packets. Both the latency and the storage requirements limit the scheme for authenticating infrequent messages broadcast by the base station. Zhu et al. have proposed a one-way key chain scheme for one-hop broadcast authentication [89]. The mechanism is known as LEAP. In this scheme, every node generates a one-way key chain of certain length and then transmits the commitment (i.e., first key) of the key chain to each neighbor, encrypted with their pair-wise shared key. Whenever a node has a message to send, it attaches to the message the next authenticated key in the key chain. The authenticated keys are disclosed in reverse order to their generation. A receiving neighbor can verify the message based on the commitment or an authenticated key it received from the sending node more recently.

Deng et al. have proposed an *intrusion tolerant routing protocol in wireless sensor networks* (INSENS) that adopts a routing-based approach to security in WSNs [2]. It constructs routing tables in each node, bypassing malicious nodes in the network. The protocol can not totally rule out attack on nodes, but it minimizes the damage caused to the network. The computation, communication, storage, and bandwidth requirements at the nodes are reduced, but at the cost of greater computation and communication at the base station. To prevent DoS attacks, individual nodes are not allowed to broadcast to the entire network. Only the base station is allowed to broadcast, and the base station is authenticated using one-way hash function so as to prevent any possible masquerading by a malicious node. Control information pertaining to routing is authenticated by the base station in order to prevent injection of false routing data. The base station computes and disseminates routing tables, since it does not have computational and energy constraints. Even if an intruder takes over a node and does not forward packets, INSENS uses redundant multipath routing, so that the destination can still reach without passing through the malicious node.

INSENS has two phases: *route discovery* and *data forwarding*. During the route discovery phase, the base station sends a request message to all nodes in the network by multi-hop forwarding. Any node receiving a request message records the identity of the sender and sends the message to all its immediate neighbors if it has not already done so. Subsequent request messages are used to identify the senders as neighbors, but repeated flooding is not performed. The nodes respond with their local topology by sending feedback messages. The integrity of the messages is protected using encryption by a shared key mechanism. A malicious node can inflict damage only by not forwarding packets, but the messages are sent through different neighbors, so it is likely that a message reaches a node by at least one path. Hence, the effect of malicious nodes is not totally eliminated, but it is restricted to only a few downstream nodes in the worst case. Malicious nodes may also send spurious messages and cause battery drain for a few downstream nodes. Finally, the base station calculates forwarding tables for all nodes, with two independent paths for each node, and sends them to the nodes. The second phase of data forwarding takes place based on the forwarding tables computed by the base station.

SPINS is a suite of security protocols optimized for sensor networks [35]. SPINS includes two building blocks: (i) *secure network encryption protocol* (SNEP) and (ii) micro version of timed efficient streaming loss-tolerant authentication protocol ($\mu TESLA$). SNEP provides data confidentiality, two-party data authentication, and data freshness for peer-to-peer communication (node to base station). $\mu TESLA$ provides authenticated broadcast as discussed already.

SPINS assumes that each node is pre-distributed with a master key $K$ which is shared with the base station at its time of creation. All the other keys, including a key $K_{encr}$ for encryption, a key $K_{mac}$ for MAC generation, and a key $K_{rand}$ for random number generation are derived from the master key using a string one-way function. SPINS uses RC5 protocol for confidentiality. If $A$ wants to send a message to base station $B$, the complete message $A$ sends to $B$ is:

$$A \rightarrow B : D_{<K_{encr}C>}, MAC(K_{mac}, C \mid D)_{<K_{encr}C>} \quad (3)$$

In the above expression, $D$ is the transmitted data and $C$ is a shared counter between the sender and the receiver for the block cipher in counter mode. The counter $C$ is incremented after each message is sent and received by the sender and the receiver respectively. SNEP also provides a counter exchange protocol to synchronize the counter value in both sides.

SNEP provides the flowing properties:

- *Semantic security*: the counter value is incremented after each message and thus the same message is encrypted differently each time.
- *Data authentication*: a receiver can be assured that the message originated from the claimed sender if the MAC verification produces positive results.
- *Replay protection*: the counter value in the MAC prevents replaying old messages by an adversary.
- *Weak freshness*: SPINS identifies two types of freshness. Weak freshness provides partial message ordering and carries no delay information. Strong freshness provides a total order on a request-response pair and allows delay estimation. IN SNEP, the counter maintains a message ordering in the receiver side and yields weak freshness. SNEP guarantees weak freshness only, since there is no guarantee to node A that a message was created by node $B$ in response to an event in node $A$.
- *Low communication overhead*: the counter state is kept at each endpoint and need not be sent in each message.

Inspired by the work on public key cryptography [60 - 62, 72], Du et al. have investigated the public key authentication problem [116]. The use of public key cryptography eases many problems in secure routing, for example, authentication and integrity. However, before a node *A* uses the public key from another node *B*, *A* must verify that the public key is actually *B*'s, i.e., *A* must authenticate *B*'s public key; otherwise, a *man-in-the-middle* attacks are possible. In general networks, public key authentication involves a signature verification on a certificate signed by a trusted third party *certificate authority* (CA) [120]. However, the signature verification operations are very expensive operations for sensor nodes. Du et al. have proposed an efficient alternative that uses only one-way hash function for the public key authentication. The proposed scheme can be divided into two stages. In the pre-distribution stage, A *Merkle tree R* is constructed with each leaf $L_i$ corresponding to a sensor node. Let $pk_i$ represent node *i*'s public key, *V* be an internal tree node, and $V_{left}$ and $V_{right}$ be *V*'s two children. The value of an internal tree node is denoted by $\Phi$. The Merkle tree can then be constructed as follows:

$$\Phi(L_i) = h(id_i, pk_i) \text{ for } i = 1,....N$$
$$\Phi(V) = h(\Phi(V_{left}) \| \Phi(V_{right})) \quad (4)$$

In the above expressions, "‖" represents the concatenation of two strings and *h* is a one-way hash function such as MD5 or SHA-1. Let *R* be the root of the tree. Each sensor node *v* needs to store the root value $\Phi(R)$ and the sibling node values $\lambda_1,.......\lambda_H$ along the path from *v* to *R*. If node *A* wants to authenticate *B*'s public key, *B* sends its public key *pk* along with the value of $\lambda_1,.......\lambda_H$ to node *A*. Then, *A* can use the same procedure to reconstruct the Merkle tree *R`* and calculate the root value $\Phi(R`)$. *A* will trust *B* to be authentic if $\Phi(R`) = \Phi(R)$. A sensor node only needs $H + 1$ storage units for the extra hash values. Based on this scheme, Du et al further extended the idea to reduce the height of the Merkle tree to improve the communication overhead of the scheme. The proposed scheme is more efficient than signature verification on certificates. However, the scheme requires that some hash values be distributed in a pre-distribution stage. This results in some scalability issues when new sensors are added to an existing WSN.

Tanachaiwiwat et al. have presented a novel secure routing protocol- *trusted routing for location aware sensor networks* (TRANS) [5]. It is primarily meant for use in data centric networks. It makes use of an asymmetric cryptographic scheme that relies on a loose-time synchronization mechanism. To ensure message confidentiality. The authors have used *μTESLA* to ensure message authentication and confidentiality. Using *μTESLA*, TRANS is able to ensure that a message is sent along a path of trusted nodes utilizing location aware routing. The base station broadcasts an encrypted message to all its neighbors. Only the trusted neighbors will possess the shared key necessary to decrypt the message. The trusted neighbors then add their locations (for the return trip), encrypt the new message with their shared key and forward the message to their neighbors closest to the destination. Once the message reaches the destination, the recipient is able to authenticate the source (base station) using the MAC corresponding to the base station. To acknowledge or reply to the message, the destination node can simply forward a return message along the same trusted path from the message was received [5].

Sen et al. have proposed a routing protocol that is resilient against packet dropping attack by malicious nodes in a WSN [121]. It essentially utilizes a single-path routing concept and hence saves energy compared to the multi-path routing protocols. If a malicious node is detected in the next-hop on the routing path, the node is efficiently bypassed and the packets are routed around the node to the base station still in a single-path. The protocol is based on a robust

*neighborhood monitoring system* (NMS) that works on promiscuous monitoring of the neighborhood of a node and detection of any possible malicious packet dropping attack by a cooperative algorithm using *neighbor list checking*.

One particular challenge to secure routing in wireless sensor networks is that it is very easy for a single node to disrupt the routing process by disrupting the route discovery process. Papadimitratos et al. have proposed a secure route discovery protocol that guarantees correct topology discovery in an ad hoc sensor network [4]. The protocol relies on the MAC and an accumulation of the node identities along the route traversed by a message. In this way, a source node discovers the sensor network topology as each node along the route from source to destination appends its identity to the message. In order to ensure that the message has not been tampered with, a MAC is constructed and can be verified both at the destination and the source (for the return message from the destination).

**5.5 Defense against Sybil attacks**

Any defense mechanism against the Sybil attack must ensure that a framework must be in place in the network to validate that a particular identity is the only identity being held by a given physical node [47]. Newsome et al. have described three orthogonal dimensions of the Sybil attack taxonomy [47]. The three dimensions are: (i) direct vs. indirect communication, (ii) fabricated vs. stolen identities, and (iii) simultaneity. In direct communication, the Sybil nodes communicate directly with legitimate nodes. In this attack, when a legitimate node sends a radio message to a Sybil node, one of the malicious devices listens to the message. In indirect communication, no legitimate nodes are able to communicate directly with the Sybil nodes. Messages sent to a Sybil node are routed through one or more of malicious nodes which pretend to pass the message on to the Sybil node. In case of fabricated identities, the attacker creates arbitrary new Sybil identities. However, if a mechanism is in place to detect false identities, an attacker cannot fabricate new identities. In this case, the attacker needs to assign other legitimate identities to Sybil nodes. This identity theft may go undetected if the attacker destroys or temporarily disables the impersonated nodes. In case of simultaneous attacks, the attacker tries to have all the Sybil identities participate in the network simultaneously. Alternatively the attacker may present a large number of identities over a period of time, while deploying a small number of identities at a given point of time. Newsome et al. primarily describe direct validation techniques, including a radio resource test. In the radio test, a node assigns each of its neighbors a different channel and listens to each of them. If the node detects a transmission on the channel, it is assumed that the node transmitting on the channel is a physical node. Similarly, if the node does not detect a transmission on the specified channel, the node assumes that the identity assigned to the channel is not a physical identity.

Another technique to defend against the Sybil attack is to use *random key pre-distribution* techniques [37 - 38, 97]. In random key pre-distribution, a random set of keys or key-related information are assigned to each sensor nodes, so that in the key set-up phase, each node can discover or compute the common keys it shares with its neighbors. The common keys are used as shared secret session keys to ensure node-to-node secrecy. Newsome et al. propose that the identity of each node is associated with the keys assigned to the node [47]. With a limited set of captured keys, there is a little probability that an arbitrarily generated identity will work.

**5.6 Detection of node replication attack**

Parno et al. have proposed a mechanism for distributed detection of node replication attacks in WSNs [51]. To address the fundamental limitations of currently existing mechanisms, e.g., single point of failure in centralized schemes, or neighborhood voting protocols that fail to detect distributed replications, the authors have proposed two algorithms that work through the collective actions of multiple nodes in a WSN. The algorithms are: (i) randomized multicast and (ii) line-selected multicast. The randomized multicast algorithm distributes location information of a node to randomly-selected witnesses, exploiting birthday paradox to detect replicated nodes. The line-selected multicast uses the network topology to detect replication as discussed below.

The randomized broadcast has evolved from traditional node-to-node broadcasting. In traditional node-to-node broadcasting, each node in the network uses an authenticated broadcast message to flood the network with its location information. Each node stores the location information of its neighbors and if it receives a conflicting claim, it revokes the offending node. This protocol can achieve 100% detection of all duplicate location claims if the broadcasts reach all the nodes. However, the total communication cost for the protocol is $O(n^2)$, which is too high for a large WSN. To reduce the communication cost of node-to-node-broadcast, deterministic multicast mechanism may be applied where a node's location claim is shared with a limited subset of deterministically chosen *witness nodes*. The witnesses are chosen as a function of the node's ID. If the adversary replicates a node, the witnesses will receive two different location claims for the same node ID. The conflicting location claims trigger the revocation of the replicated node. The randomized multicast approach suggested by Parno et al. improves the robustness of the deterministic multicast. It randomizes the witnesses for a given node's location claim, so that the adversary cannot anticipate their identities. When a node announces its location, each of its neighbors sends a copy of the location claim to a set of randomly selected witness nodes. If the adversary replicates a node, then two sets of witnesses will be selected. In a network of *n* nodes, if each location produces $\sqrt{n}$ witnesses, then, the birthday paradox predicts at least one collision with high probability, i.e., at least one witness will receive a pair of conflicting location claims. The two conflicting location claims form sufficient evidence to revoke the node, so the witness can flood the pair of location claims through the network, and each node can independently confirm the revocation decision. Unfortunately, however, the communication and storage overheads for randomized multicast is too high- $O(n^2)$ and $O(\sqrt{n})$ respectively. The authors have suggested some enhancements for improving the communication and storage overhead.

To reduce the communication cost of the randomized multicast approach, Parno et al. have proposed an alternative algorithm- the line selected multicast. It is based on the *rumor routing protocol* [122]. The idea is that a location claim traveling from source s to destination d will also travel through several intermediate nodes. If each of these nodes records the location claims, then the path of the location claim through the network can be thought of as a line segment. The destination of the location claim is one of the randomly chosen witnesses. As the location claim routes through the network towards a witness node, the intermediate sensors check the claim. If a conflicting location claim ever crosses a line segment, then the node at the intersection detects the conflict and initiates a revocation broadcast. The line selected multicast algorithm has communication overhead of $O(n\sqrt{n})$ as long as each line segment is of length $O(\sqrt{n})$ nodes. The storage overhead of the algorithm is $O(\sqrt{n})$.

### 5.7 Defense against traffic analysis attack

Deng et al. have proposed a mechanism for defending against traffic analysis attack in a WSN [55]. The author have argued that since the base station is a central point of failure, once the location of the base station is discovered, an adversary can disable or destroy the base station, thereby rendering the data-gathering functionalities of the entire WSN ineffective. Two classes of traffic analysis attacks in WSNs are identified: (i) *rate monitoring attack*, and (ii) *time correlation attack*. In time correlation attack, an adversary monitors the packet sending rate of nodes near the adversary, and moves closer to the nodes that have a higher packet sending rate. In a time correlation attack, an adversary observes the correlation in sending time between a node and its neighbor node that is assumed to be forwarding the same packet, and deduces the path by following the sound of each forwarding operation as the packet propagates towards the base station. The mechanism proposed by the authors prevents rate monitoring attack and time correlation attack. The mechanism involves four techniques. First, a multiple parent routing scheme is introduced that allows a sensor node to forward a packet to one of its multiple parents. This makes the patterns less pronounced in terms of routing packets towards the base station. Second, a controlled random walk is introduced into the multi-hop path traversed by a packet through the WSN towards the base station. This distributes packet traffic, thereby rendering the rate monitoring attack less effective. Third, random fake paths are introduced to confuse an adversary from tracking a packet as it moves towards a base station. This mitigates the effectiveness of time correlation attacks. Finally, multiple, random areas of high communication activities are created to deceive an adversary as to the true location of the base station, which further increases the difficulty of rate monitoring attacks. The combination of these four strategies makes the proposed mechanism extremely robust to any traffic analysis attack.

**5.8 Defense against attacks on senor privacy**

The attacks on information privacy in WSNs have been discussed in Section 4. In this subsection some schemes protecting information privacy in WSNs are discussed.

**5.8.1 Anonymity mechanisms**

Precise location information enables accurate identification of a user. This is a serious threat to privacy. One way to handle this problem is to make data source anonymous. An anonymity mechanism depersonalizes the data before it is released from the source. Gruteser et al. have presented an analysis on the feasibility of anonymizing location information in location-based services in an automotive telematics environment [123]. Beresford et al. have proposed anonymity techniques for an indoor location system based on the *Active Bat* [124]. In [125], an efficient and reliable routing protocol for wireless ad hoc and mesh networks has presented that ensures anonymity of the user. The user anonymity, authentication and data privacy is achieved by application of a novel protocol that is based on Rivest's ring signature scheme [126].

Since ensuring total anonymity is almost an impossible proposition, in almost all practical scenarios, a tradeoff is to be made between anonymity and disclosure of public information in most of the privacy protection mechanisms. Four approaches have been proposed by researchers in this direction [52, 127 - 129] for WSNs. These approaches are: (i) decentralization of storage of sensitive data, (ii) establishment of secure channel for communication, (iii) changing the pattern of data traffic, and (iv) exploiting mobility of the nodes. The sensitive location data is to be stored in a spanning tree of nodes so that no single node holds a complete view of the location information. Communication using secure protocols such as SPINS [35] will make eavesdropping and active attack on a WSN extremely difficult. The data traffic pattern may be changed by selectively inserting some bogus data in the network traffic so that traffic analysis by an external entity will not be successful. Mobile sensor nodes make attack on location privacy very difficult. The *Cricket* system [128] is a location-support system for mobile, and location-dependent applications inside large buildings. It allows applications running on mobile and static nodes to learn their physical location from a set of listeners. The listeners hear and analyze information from beacons in a building. The location sensors are placed on the mobile devices instead of some static locations in the building, and the location information is not disclosed during the position determination process.

**5.8.2 Policy-based approaches**

In policy-based defense mechanisms the access control decisions and authentication techniques are made on the basis of a specified set of privacy policies. Molnar et al. have presented the concept of private authentication and demonstrated its application in *radio frequency identification* (RFID) domain [130]. Duri et al. propose a policy-based framework for protecting sensor information, where a computer in side a car acts as a trusted agent for location privacy [131]. Snekkenes introduces various parameters for access control that enable specifying policies in the context of a mobile network [132]. Some of the parameters are: time of request, location, speed, and identity of the located object. Myles et al. describe the architecture of a centralized location server that controls access requests from client applications through a set of validator modules based on a set of XML-coded privacy policies [133]. Hengartner et al. have discussed various challenges that arise for the specification and implementation of policies controlling access to location information [134]. The authors have also presented a design framework of an access control mechanism.

**5.8.3 Information flooding**

Ozturk et al. proposed various modifications to WSN routing protocols for protecting the location information of a source node [53]. In particular, the authors have discussed a set of flooding protocols. The randomized data routing and phantom traffic generation mechanism are used so that it is difficult for an adversary to track any data source. For ensuring privacy of source location the authors have discussed four types of flooding-based routing protocols: (i) baseline flooding, (ii) probabilistic flooding, (iii) flooding with fake messages, and (iv) phantom flooding. They are described as follows:

- *Baseline flooding*: In the baseline flooding, every node in the network forwards a message only once, and no node retransmits a message that it has previously transmitted. When a message reaches an intermediate node, the node first checks whether it has received and forwarded the message before. If this is its first time, the node broadcasts the message to all its neighbors. Otherwise, it just discards the message.
- *Probabilistic flooding*: In probabilistic flooding, only a subset of nodes in the entire network participates in data forwarding, while the others simply discard the messages they receive. One possible weakness of this approach is that some messages may get lost in the network and as a result affect the overall network connectivity. However, Ozturk et al. [53] have proved analytically that this is not a significant problem.
- *Flooding with fake messages*: Flooding cannot provide privacy protection because an adversary can easily identify the shortest path between a source and a sink, and use it to backtrack to the source location. The main reason for lack of location privacy is that there is only one source node. One approach that can alleviate the risk of source-location privacy breaching is to augment the flooding protocols so that more sources can be introduced that inject fake messages into the network. If the fake messages have the same length as the real messages and they are also encrypted, it will be impossible for an adversary to distinguish between them.
- *Phantom flooding*: Phantom flooding has the same principle as that of probabilistic flooding. It too attempts to direct messages to different locations of the network so that the adversary cannot receive a steady stream of messages to track the source. However, probabilistic flooding is not very effective since shorter paths are more likely to deliver more messages. Phantom flooding entices an attacker away for the real source and towards a fake source, called the phantom source. In phantom flooding, every message experiences two phase: (1) a walking phase, which may be a random walk or a directed walk, and (2) a subsequent flooding meant to deliver the message to the sink. When the source sends out a message, the message is unicast in a random fashion within the first $h_{walk}$ hops. This is called the *random walk* phase. After the $h_{walk}$ hops, the message is flooded using the baseline flooding technique. This is the *flooding* phase. Phantom flooding significantly improves the privacy and network safety period because every message may take a different (shortest) path to reach any node in the network.

Deng et al. have addressed the problem of defending a base station against physical attacks by concealing the geographic location of the base station [55]. The authors have investigated several countermeasures against traffic analysis techniques aimed at disguising the location of a base station. In the proposed mechanism, a degree of randomness is introduced while selecting the multi-hop route to the base station. Then, random fake packets are introduced as a packet moves towards a base station. The metrics such as total entropy of the network, total energy consumed, and the ability to guard against heuristic-based techniques to locate the base station are evaluated analytically as well as by extensive simulations.

Xi et al. [135] have described a successful attack on the *flooding-based phantom routing* proposed by Ozturk et al [53]. The authors have also proposed *greedy random walk* (GROW) protocol, a two-way random walk, i.e., from both source and sink, to reduce the chance an eavesdropper can collect the location information. In the proposed mechanism, the sink first initiates an *N*-hop random walk, and the source then initiates an *M*-hop random walk. Once the source packet reaches an intersection of these two paths, it is forwarded through the path created by the sink. Local broadcasting is used to detect when the two paths intersect. In order to minimize the chance of backtracking along the random walk, the nodes are stored in a bloom filter as the walk progresses. At each stage, the intermediate nodes are checked against the bloom filter to ensure that backtracking is minimized.

**5.9 Intrusion detection**

The security mechanisms implemented in secure routing protocols and secure data aggregation protocols are configured beforehand to prevent an attacker from breaking the security of the network. However, these security mechanisms alone cannot ensure security of a WSN. Since it is possible for an attacker to compromise a sensor node, it is easy for him to inject false data into a WSN. Authentication and data encryption are not enough for ensuring data security. Another approach to protect WSNs involves mechanisms for detecting and reacting to intrusions.

An *intrusion detection system* (IDS) monitors a host or network for suspicious activity patterns outside normal and expected behavior [29]. It is based on the assumption that there exists a noticeable difference in the behavior of an intruder and legitimate user in the network such that an IDS can match it with pre-programmed or possible learned rules. Based on the analysis model used for analyzing the audit data to detect intrusions, intrusion detection systems are usually classified into two types: (i) rule-based intrusion detection systems and (ii) anomaly-based intrusion detection systems [136]. Rule-based intrusion detection systems are used to detect known patterns of intrusions as in [137] and [138]. The anomaly-based systems are used to detect new or unknown intrusions as in [139] and [140] rule-based IDS has a low false-alarm rate compared to an anomaly-based system, and an anomaly-based IDS has a high intrusion detection rate in comparison to a rule-based system. Distributed IDSs have higher detection efficiency. Sen et al. have presented the model of a distributed IDS that consists of a large number of autonomous and cooperating agents [141]. The architecture exploits inter-agent communication and distributed computation to achieve a high detection efficiency with very low rates of false positives and false negatives.

However, WSNs are generally application-specific and lack basic information on topology, normal usage, expected communication patterns, etc. It is impractical to pre-install some fixed patterns in sensors before they are deployed. Moreover, due to constraints in sensors, to learn and detect these parameters after deployment is both time and energy consuming. Thus, existing intrusion detection schemes in ad hoc networks may not be adapted to WSNs.

The research on intrusion detection in WSNs is still preliminary. Current research focuses on how to detect and eliminate injected false information. Thus, cooperation among sensors, especially neighboring nodes, is necessary to decide the validity of a report. The following subsection discusses some existing mechanisms of intrusion detection for WSNs.

### 5.9.1 Intrusion detection in WSNs

Brutch et al. have discussed various types of possible attacks against WSNs and presented three different architectures for intrusion detection [142]. The first is a stand-alone architecture. In this case, each node functions as an independent intrusion detection system and is responsible for detecting attacks directed towards it. The nodes do not exchange and intrusion data and no cooperative detection mechanisms are deployed. The second architecture is a distributed and cooperative architecture. In this architecture, an intrusion detection agent is deployed on each node. While the local agents are responsible for detecting local attacks on the nodes, they also cooperate among themselves by exchanging intrusion related data to detect global intrusion attempts. The third architecture proposed by the author is a hierarchical architecture. This is suitable for a multi-layered WSN, where the network is divided into clusters with the cluster-head node being responsible for routing within a cluster. The multi-layered networks are primarily used for event correlation.

Zhu et al. proposed an *interleaved hop-by-hop* authentication (IHOP) scheme in [143]. IHOP guarantees that the base station will detect any injected false data packets when no more than a certain number $t$ of nodes are compromised. The sensor network is organized in a cluster-based hierarchy. Each cluster-head builds a route to the base station and each intermediate node has an upper associate node and a lower associate node that is $t + 1$ hops away. IHOP uses a number of shared keys: (i) every node shares a master key with the base station, (ii) each node knows its one-hop neighbors and has established a pair-wise key with each of them, (iii) a node can establish a pair-wise key with another node that is multiple hops away if needed.

Further, IHOP also assumes that the base station has a mechanism to authenticate broadcast messages, e.g., *μTESLA*. A cluster-head collects information from its members and sends a report to the base station only when at least $t + 1$ sensors observer the same result. Meanwhile a cluster-head also collects the MACs from detecting nodes. Each detecting node sends two MACs to the cluster-head: a MAC using the key shared with the base station, referred to as the individual MAC, and a MAC using the key shared with its upper associate nodes, referred to as the pair-wise MAC. The cluster-head then compresses the $t + 1$ individual MACs by XORing them to reduce the size of the report. However, the pair-wise MACs are not compressed for transmission. If they were, a node replaying the message would not be able to extract the pair-wise MACs and a compressed MAC for the base station. When an intermediate node receives a report, it verifies the MAC of its lower associate node. If it fails, the report is eliminated. Otherwise, it removes the MAC, generates a new MAC using its upper associate node pair-wise key, and appends it to the report. However, the pair-wise MACs are not compromised for transmission. If they were, a node relaying the message would not be able to extract the pair-wise MACs of interest to it. Thus, a legitimate report includes $t + 1$ pair-wise MACs and a compressed MAC for the base station. When an intermediate node receives a report, it verifies the MAC of its lower associate node. If it fails, the report is eliminated. Otherwise, it removes the MAC, generates a new MAC using its upper associate node pair-wise key, and appends it to the report.

IHOP ensures that the base station can detect false data packets when no more than $t$ nodes are compromised. However, the authors [143] have not shown how to select the parameter $t$ for a sensor network.

Wang et al. proposed a scheme to detect whether a node is faulty or malicious with the collaboration of neighbor nodes [144]. In the proposed scheme, when a node suspects that one of its neighbors is faulty, it sends out messages to request the opinions on the behavior of this suspected node from other neighbors of the suspect. After collecting the results, the node analyzes the results to diagnose whether the suspect has a fault. The authors formalized the problem as how to construct a dominating tree to cover all the neighbors of the suspect and further proposed two tree-based propagation collection protocols to construct a dominating tree and collect information via the tree structure.

Albers et al. have proposed an intrusion detection architecture based on *local intrusion detection system* (LIDS) on each node in a wireless ad hoc network [145]. In order to detect a network-wide intrusion, the LIDS on the nodes collaborate with each other and exchange two types of data- security data and intrusion alerts. The security data is used to exchange information with other network hosts. The intrusion alerts are used to inform LIDS in the neighboring nodes to exchange intrusion related information. Although the framework is for an ad hoc network, its approach of local anomaly detection and cooperatively detecting any network-wide intrusion can be adapted to develop an intrusion detection mechanism for a WSN [6].

Intrusion detection in WSNs is still largely open to research. Key research issues include the following:

- Due to the constraints in WSNs, intrusion detection has many aspects not of concern in other network types. The problem of intrusion detection needs to be well defined in WSNs.
- The proposed IDs protocols in the literature focus on filtering injected false information only [2, 11, 143]. These protocols need to be improved to address scalability issues.

**5.10 Secure data aggregation**

Data communication constitutes an important share of the total energy consumption of a sensor network. Simulation [35] shows that data transmission accounts for 71 percent of the energy cost of computation and communication for the *secure network encryption protocol* (SNEP) protocols. An efficient data aggregation mechanism can greatly help in optimizing the energy consumption.

In a WSN, there are certain nodes called aggregators which are responsible to carry out data aggregation operations. If an aggregator node is compromised, it is easy for an adversary to inject false data into the network. Another possible attack is to compromise a sensor node and inject forged data through it. Without authentication, the attackers may fool the aggregators into reporting false data to the base station. Secure data aggregation requires authentication, confidentiality, and integrity of data to be maintained. Moreover, secure data aggregation also requires cooperation among the sensor nodes to identify the compromised sensors.

Before discussing some secure aggregation mechanisms existing in the literature, an overview of some well-known aggregation techniques are presented.

In [6], the authors have proposed a clustering-based algorithm that uses directed diffusion technique to gather a global perspective utilizing only the local nodes in each cluster. The nodes are assigned different level with level 0 being assigned to the nodes lying at the lowest level. While the nodes at the highest level can communicate across the clusters, the nodes at the lower levels communicate among each other in the same cluster via the cluster head node. This effectively enables localized cluster computation while the higher level nodes communicate the local information of the clusters to achieve a global picture.

In [8], the authors have proposed a mechanism called *tiny aggregation* (TAG) service. It is a generic data aggregation mechanism that involves a language similar to SQL to generate queries in a WSN. The base station generates a query using this language. The sensor nodes sends the reply using routes constructed based on a routing tree. At each point in the tree, the data is aggregated using some aggregation function that was defined in the initial query sent.

Srivastava et al. have proposed a summary structure for supporting fairly complex aggregate functions, such as median and range queries [10]. In addition, computation of relatively easier functions such as min/max, sum, and average are also supported in the proposed framework. However, more complex aggregate, such as the most frequently reported data value is not supported. The computed aggregate functions are approximate but the estimated errors are statistically bounded.

A number of secure aggregation protocols for WSNs exist in the literature. However, fundamentally there is a conflict of interests in data confidentiality and data aggregation. Confidentiality requires the data to be transmitted in ciphertext mode, whereas data aggregation is usually done on plaintext contents. A straight-forward method is to invoke end-to-end encryption before executing data aggregation. This strategy, however, has a shortcoming. The encryption and decryption operations involve a substantial computation overhead. An alternative method is to provide data aggregation on concealed data, which requires a particular class of encryption transformation. However, this method usually reduces the security level [103].

**Figure 5** shows a taxonomy of secure data aggregation protocols for WSNs. There are two categories of secure aggregation protocols: (i) plaintext-based, and (ii) ciphertext-based.

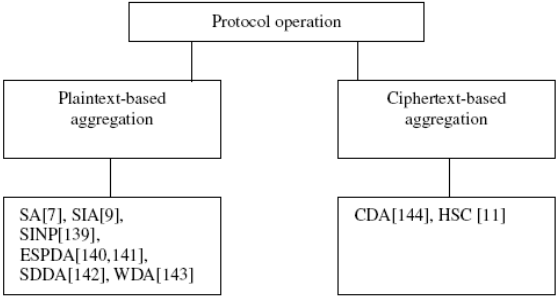

**Fig. 5.** Secure data aggergation in WSNs; a taxonomy [171]
(Source: Y. Wang, G. Attebury, and B. Ramamurthy, IEEE Communications Surveys and Tutorials, Vol. 8, no. 2, pp. 2-23, 2006)

### 5.10.1 Secure aggregation on plaintext data

Hu et al. have proposed a *secure aggregation* (SA) protocol that uses the *µTESLA* protocol [7]. The protocol is resilient to both intruder devices and single device key compromises. In the proposition, the sensor nodes are organized into a tree where the internal nodes act as the aggregators. However, the protocol is vulnerable if a parent and one of its child nodes are compromised, since due to the delayed disclosure of symmetric keys, the parent node will not be able to immediately verify the authenticity of the data sent by its children nodes.

Przydatek et al. have presented a *secure information aggregation* (SIA) framework for sensor networks [9]. The framework consists of three categories of node: a home server, base station and sensor nodes. A base station is a resource-enhanced node which is used as an intermediary between the home server and the sensor nodes, and it is also the candidate to perform the aggregation task. SIA assumes that each sensor has a unique identifier and shares a separate secret cryptographic key with both the home server and the aggregator. The keys enable message authentication and encryption if data confidentiality is required. Moreover, it further assumes that the home server and the base station can use a mechanism, such as *µTESLA*, to broadcast authenticated messages. The proposed solution follows *aggregate-commit-prove* approach.

In the first phase- aggregate- the aggregator collects data from sensors and locally computes the aggregation result using some specific aggregate function. Each sensor shares a key with the aggregator. This allows the aggregator to verify whether the sensor reading is authentic. However, there is a possibility that a sensor may have been compromised and an adversary has captured the key. In the proposed scheme there is no mechanism to detect such an event.

In the second phase, i.e. commit phase, the aggregator commits to the collected dat. This phase ensures that the aggregator actually uses the data collected from the sensors, and the statement to be verified by the home server about the correctness of computed results is meaningful. One efficient mechanism for committing is a *Merkle hash-tree* construction [146]. In this method, the data collected from the sensors is placed at the leaves of a tree. The aggregator then computes a binary hash tree staring with the leaf nodes. Each internal node in the hash tree is computed as the hash value of the concatenation of its two children nodes. The root of the tree is called the commitment of the collected data. As the hash function in use is collision free, once the aggregator commits to the collected values, it cannot change any of the collected values.

In the third and final phase, the aggregator and the home server engage in a protocol in which the aggregator communicates the aggregation result. In addition, aggregator uses an interactive proof protocol to prove correctness of the reported results. This is done in two logical steps. In the first step, the home server ensures that the committed data is a good representation of the sensor data readings collected. In the second step, the home server checks the reliability of the aggregator output. This is done by checking whether the aggregation result is close to the committed results. The interactive proof protocol varies depending on the aggregation function is being used. Moreover, the authors also presented efficient protocols for secure computation of the median and the average of the measurements, for the estimation of the network size, and for finding the minimum and maximum sensor reading.

Deng et al. proposed a collection of mechanisms for *securing in-network processing* (SINP) for WSNs [147]. Security mechanism were proposed to address the downstream requirement that sensor nodes authenticate commands disseminated from parent aggregators and the upstream requirement that aggregators authenticate data produced by sensors before aggregating that data. In the downstream stage, two techniques are involved: one way functions and *μTESLA*. The upstream stage requires that a pair-wise key be shared between an aggregator and its sensor nodes.

Cam et al. proposed an *energy-efficient secure pattern-based data aggregation* (ESPDA) protocol for wireless sensor networks [148, 149]. ESPDA is applicable for hierarchy-based sensor networks. In ESPDA, a cluster-head first requests sensor nodes to send the corresponding pattern code for the sensed data. If multiple sensor nodes send the same pattern code to the cluster-head, only one of them is permitted to send the data to the cluster-head. ESPDA is secure because it does not require encrypted data to be decrypted by cluster-heads to perform data aggregation.

Cam et al. have introduced another *secure differential data aggregation* (SDDA) scheme based on pattern codes [150]. SDDA prevents redundant data transmission from sensor nodes by implementing the following schemes: (1) SDDA transmits differential data rather than raw data, (2) SDDA performs data aggregation on pattern codes representing the main characteristics of the sensed data, and (3) SDDA employs a sleep protocol to coordinate the activation of sensing units in such a way that only one of the sensor nodes capable of sensing the data is activated at a given time. In the SDDA data transmission scheme, the raw data from sensor nodes is compared to reference data with the difference data being transmitted. The reference data is obtained by taking the average of previously transmitted data.

Du et al. proposed a *witness-based data aggregation* (WDA) scheme for WSNs to assure the validation of the data fusion nodes to the base station [151]. To prove the validity of the fusion results, the fusion node has to provide proofs from several witnesses. A witness is one who also conducts data fusion like a data fusion node, but does not forward its result to the base station. Instead, each witness computes the MAC of the result and then provides it to the data fusion node, which must forward the proofs to the base station.

Wagner studied secure data aggregation in sensor networks and proposed a mathematical framework for formally evaluating their security [152]. The robustness of an aggregation operator against malicious data is quantified. Ye et al. propose a *statistical en-route filtering mechanism* to detect any forged data being sent from the sensor nodes to the base station of a WSN using multiple MACs along the path from the aggregator to the base station [11].

**5.10.2 Secure aggregation on ciphertext data**

Secure aggregation of ciphertext data in WSNs is required to preserve privacy of the sensor nodes. Efficient in-network data aggregation with preservation of data privacy is an important requirement in many WSN applications [153-155, 157, 162]. As a key approach to fulfilling this requirement of private data aggregation of sensor nodes, *concealed data aggregation* (CDA) schemes have been proposed in which multiple source nodes send encrypted data to a sink along a converge-cast tree with aggregation of ciphertext being performed over the route [153 - 155, 157, 158]. Two ciphertext-based secure data aggregation schemes have been proposed in [154] and [155]. The propositions are based on a particular encryption transformation: a *privacy homomorphism* (PH). A privacy homomorphism is an encryption transformation that allows direct computation on the encrypted data. Let $Q$ and $R$ denote two rings, and let + denote addition and x denote multiplication on the two rings. Let $K$ be the key space. We denote an encryption transformation $E : K \times Q \rightarrow R$ and the corresponding decryption transformation $D : K \times R \rightarrow Q$. Given $a, b \in Q$ and $k \in K$, the following operation is termed as *additively homomorphic*.

$$a + b = D_k(E_k(a) + E_k(b)) \quad (5)$$

Similarly, the following operation is termed as multiplicatively homomorphic [159].

$$a \times b = D_k(E_k(a) \times E_k(b)) \qquad (6)$$

The *concealed data aggregation* (CDA) scheme proposed in [155] is based on the PH proposed in [160]. Although the study in [161] has shown that the proposed PH in [160] is insecure against chosen plaintext attacks for some parameter settings, the authors in [155] claimed that for the WSN data aggregation scenario, the security level is still adequate and the proposed PH method in [160] can be employed for encryption. CDA can be used to calculate SUM and AVERAGE in a hierarchical WSN. To calculate AVERAGE, an aggregator needs to know the number of sensor nodes $n$.

Castelluccia et al. proposed a simple and provable secure additively *homomorphic stream cipher* (HSC) that allows for the efficient aggregation of encrypted data [154]. The new cipher uses modular addition and is therefore very well suited for CPU-constrained devices such as those in WSNs. The aggregation based on this cipher can be used to efficiently compute statistical values such as the mean, variance, and standard deviation of sensed data while achieving significant bandwidth gain.

Secure data aggregation activity is an extremely important issue in WSNs. Several secure data aggregation protocols have been proposed by the researchers. However, no comparisons have been conducted on these proposed protocols. Further evaluations are required to get an idea about the performance of these protocols. The performance metrics might include security, processing overhead, communication overhead, energy consumption, data compression etc. Moreover, new data aggregation protocols are needed to address higher scalability and higher reliability against aggregator and sensor node cheating [103].

## 5.11 Defense against physical attacks

To protect against a possible physical attack, sensor nodes may be equipped with special hardware. The sensor nodes in a WSN may be protected against tampering by tamper-proofing the physical packages of the sensors [29]. Researchers have also proposed mechanisms that focus on building tamper-resistant hardware in order to make the memory contents on the sensor chip inaccessible to a potential external attacker [20, 21, 24]. Special-purpose software and hardware may also be deployed outside the sensor nodes to detect physical tampering. Self-termination of sensor nodes is an effective mechanism to defend against possible data theft in the event of a physical attack. The basic idea in this case is that whenever a sensor senses an attack it kills itself and destroys all data and keys stored in its memory. This is particularly feasible in a large-scale WSN where there is enough redundancy of information and connectivity among the nodes. However, the main challenge is to accurately identify a physical attack. A simple solution is to periodically verify the neighborhood information for each node. In case of a mobile sensor network, this is an open problem.

A number of techniques have been discussed for extracting protected data from card processor [20 - 24]. These techniques include manual micro-probing, laser cutting, focused ion-beam manipulation, glitch attacks, power analysis etc. Most of these techniques may be used to launch physical attacks on sensor nodes in a WSN.

Anderson et al. have proposed counter-measures for each of these attacks [21]. In [20 - 24], the authors describe techniques for extracting protected software and data from smart card processors. This includes manual micro-probing, laser cutting, focused ion-beam manipulation, glitch attacks, and power analysis, most of which are also possible physical attacks on the sensor. Based on an analysis of these attacks, Andersen et al give examples of low-cost protection countermeasures that make such attacks considerably more difficult, including.

Deng et al. have proposed various approaches for protecting sensors by deploying components outside them [163]. Sastry et al. have presented ECHO protocol for secure and reliable location verification of sensor nodes in a WSN [25]. The scheme is based on the physical properties of sound and RF signal propagation from the sensor nodes. It is not possible for an adversary to cheat and falsely claim a shorter distance from the base station by transmitting its ultrasonic sound response early, because it will not be able to produce the required nonce for verification.

In [2], the authors presented defense mechanisms against *search-based physical attacks*. The authors have also discussed a systematic modeling framework for *blind physical attacks* [27]. The defense mechanism against physical attacks as proposed by the authors involves two phases. In the first phase, the sensors detect the attacker and send out attack notification messages in the network. In the second phase, the sensors that receive the notification messages schedule their states to switch off mode. Seshadri et al. have proposed a mechanism called *software-based attestation for embedded devices* (SWATT), to detect a sudden and abrupt change in the memory contents of a sensor node [26]. An abrupt change in the memory content of a sensor indicates possibility of a physical attack.

**5.12 Trust management**

Application of trust and reputation-based frameworks for enforcing high-level of security in WSNs is another approach. In fact trust-based schemes can protect against attacks which are beyond the capabilities of the cryptographic security. For example issues like judging the quality and reliability of sensor nodes and wireless links, data aggregation reliability and correctness of aggregator nodes, timeliness in packet forwarding of the sensors etc are can be addressed effectively in a systematic manner with the help of a trust-based framework. However, trust-based models usually involve high computational overhead, and building an efficient scheme for resource-constrained WSNs is a very challenging task. A comprehensive discussion on basic concepts of trust and reputation and various security mechanisms based on these concepts for WSNs are presented in [164].

Pirzada et al. [165] have proposed an approach for building trust relationship between the nodes in an ad hoc network. It is assumed that the nodes in the network passively monitor the packets received and forwarded by the other nodes. The receiving and forwarding activities by the nodes are termed as *events*. Events are observed and given a weight, depending on the type of application requiring a trust relationship with other nodes. The weights reflect the significance of the observed events for the corresponding application. The trust values for all events from a node are combined using weights to compute an aggregate trust level for the node. The computed trust values are used as link weights for the computation of routes. Links which connect more trust-worthy nodes will be having smaller weights. A shortest-path routing algorithm would compute the most trustworthy paths in a network.

In [166], the author has described methods of finding paths from a source node to a designated target node in a peer-to-peer computing paradigm. Extending this approach, Zhu et al. [19] provide a practical approach to compute trust in wireless networks by treating individual mobile device as a node of a delegation graph $G$ and mapping a delegation path from a source node $S$ to a target node $T$ into an edge in the corresponding transitive closure of the graph $G$. From the edges of the transitive closure of the graph $G$, the trust values of the wireless links are computed. In the proposed trust-based framework, an undirected transitive signature scheme is used within the authenticated transitive graphs.

In [167], a secure and efficient searching scheme for peer-to-peer networks has been proposed that utilizes topology adaptation by contsruting an overlay of trusted peers where the neighbors are selceted based on their trust ratings and content similarities. Using a robust trust management scheme, the scheme provides a highly reliable framework for protecting the privacy os users and data in the network.

Yan et al. have proposed a security solution based on trust framework to ensure data protection, secure routing and other security features in an ad hoc network [18]. Mechanisms of logical and computational trust analysis and evaluation are applied on the nodes. Each node evaluates the trust of its peers based on factors such as experience statistics, data value, intrusion detection results, recommendations from its other neighbors. Ren et al. have presented a technique to establish trust relationships among nodes in an ad hoc network [16]. The proposed framework is a probabilistic solution based on a distributed trust model. A secret dealer is introduced only in the system bootstrapping phase to initiate the trust propagation. Shorter and more robust trust chains are subsequently developed among the nodes. A fully self-organized trust establishment approach is then adopted to conform to the dynamic membership changes.

Ganeriwal et al. have proposed a reputation-based framework for high integrity sensor networks [12]. The framework employs a beta distribution for reputation representation, updates, and integration. Using Beta distribution for reputation computation and exploiting statitistical theory of estimation, a secure and robust data aggregation scheme for WSNs has been presented in [168]. Tanachaiwiwat et al. [17] have proposed a mechanism of location-centric isolation of nodes exhibiting misbehavior and trust-based routing among nodes a in sensor networks. The trust valued of a node is computed based on the cryptographic suite being applied, availability statistics and the packet forwarding information of the node. If computed trust associated with a node falls below a threshold, the node's location is considered insecure and it is avoided in routing process. The robust reputation computation model allows accurate detection of node misbehavior.

In [169], a reputation- and trust-based security framework for ad hoc networks has been proposed for detecting malicious packet-dropping attacks. The mechanism is based on a trust model that computes repuation values for the nodes in a network. A similar scheme based on cooperation of nodes in a neighborhood and a distributed algorithm for reputation computation has been presented in [170].

Linag et al. have carried out extensive work on development of models and evaluating robustness and security of various aggregation algorithms in open and untrusted environment [13, 14]. These models may be adapted for deployment of trust-frameworks in WSN. In [15], the authors have proposed a model called *personalized trust* (PET) for nodes in a WSN. In [14], for aggregation of various ratings received from its peer sensor nodes, a comprehensive analytical and inference model of trust have been presented. The authors have identified two types of uncertainties in a rating system in an open computing environment: the uncertainties associated with rating aggregation algorithms and uncertainties resulting from other algorithm-independent design factors. The authors have shown that complex aggregation algorithms are not suitable in many cases due to memory limitations in the sensor nodes for storing knowledge related to computation of a trust-based framework. The simulation results show that it is better to treat ratings received from different evaluators (i.e., nodes) with equal weight and simply compute the average to arrive at the final trust value. This approach not only has a very low computational overhead, it also produces very satisfactory result in practice. The authors also observe that for a trust model most important and critical issue is how to adaptively adjust the parameters of the model based on the change in environment.

## 6. Conclusions and Future Trends

Although research efforts have been made on cryptography, key management, secure routing, secure data aggregation, and intrusion detection in WSNs, there are still some challenges to be addressed. First, the selection of the appropriate cryptographic methods depends on the processing capability of sensor nodes, indicating that there is no unified solution for all sensor networks. Instead, the security mechanisms are highly application-specific. Second, sensors are characterized by the constraints on energy, computation capability, memory, and communication bandwidth. The design of security services in WSNs must satisfy these constraints. Third, most of the current protocols assume that the sensor nodes and the base station are stationary. However, there may be situations, such as battlefield environments, where the base station and possibly the sensors need to be mobile. The mobility of sensor nodes has a great influence on sensor network topology and thus raises many issues in secure routing protocols. Following research issues for security in WSNs are particularly important:

- *Exploit the availability of private key operations on sensor nodes*: recent studies on public key cryptography have shown that public key operations may be practical in sensor nodes. However, private key operations are still very expensive to realize in sensor nodes. As public key cryptography can greatly ease the design of security in WSNs, improving the efficiency of private key operations on sensor nodes is highly desirable.
- *Secure routing protocols for mobile sensor networks*: mobility of sensor nodes has a great influence on sensor network topology and thus on the routing protocols. Mobility can be at the base station, sensor nodes, or both. Current protocols assume the sensor network is stationary. New secure routing protocols for mobile sensor networks need to be developed.
- *Time synchronization issues*: current broadcast authentication schemes such as μTESLA and its extensions require sensor networks to be loosely time synchronized. This requirement is often hard to meet and new techniques that do not have such requirements are in great demand.
- *Scalability and efficiency in broadcast authentication protocols*: novel schemes with higher scalability and efficiency need to be developed for authenticated broadcast protocols. The recent progress in public key cryptography may facilitate the design of authenticated broadcast protocols.
- *Defending DoS attacks:* defending DoS attack is a great challenge. In the simplest from of this attack, an adversary attempts to disrupt communication by transmitting a broadcast signal of high strength. The adversary can also inhibit communication violating the MAC protocol by transmitting frames while a neighbor is also by transmitting or by continuously requesting channel access with a *request-to-send* (RTS). New techniques for dealing with these attacks are needed.
- *Continuous stream security in WSNs*: current work on security in sensor networks focuses on discrete events such as temperature and humidity. Continuous stream events such as video and images are not discussed. Video and image sensors for WSNs might not be widely available now, but will be likely in the future. Substantial differences in authentication and encryption exist between discrete events and continuous events, indicating that there will be distinctions between continuous stream security and the current protocols in WSNs.
- *QoS and security*: performance is generally degraded with the addition of security

services in WSNs. Current studies on security in WSNs focus on individual topics such as key management, secure routing, secure data aggregation, and intrusion detection. QoS and security services need to be evaluated together in WSNs.

## References

1. Akyildiz, F., W. Su, Y. Sankarasubramaniam, and E. Cayirci. August 2002. "A Survey on Sensor Networks." *IEEE Communications Magazine* 40 (80): 102 – 114.
2. Deng, J., R. Han, and S. Mishra. November 2002. "INSENS: Intrusion-Tolerant Routing in Wireless Sensor Networks", Technical Report CU-CS-939-02, Department of Computer Science, University of Colorado at Boulder, November 2002.
3. Karp. B. and H. T. Kung. "GPSR: Greedy Perimeter Stateless Routing for Wireless Networks." In *Proceedings of the 6th Annual International Conference on Mobile Computing and Networking (MobiCom'00)*, 243-254, August 2000, Boston, Massachusetts, USA.
4. Papadimitratos, P., and Z.J. Haas. January 2002. "Secure Routing for Mobile Ad hoc Networks." In *Proceedings of the SCS Communication Networks and Distributed System Modeling and Simulation Conference (CNDS 2002)*, 27-31, San Antonio, TX, USA.
5. Tanachaiwiwat, S., P. Dave, R. Bhindwale, and A. Helmy. November 2003. "Routing on Trust and Isolating Compromised Sensors in Location-Aware Sensor Networks." In *Proceedings of the 1st International Conference on Embedded Networked Sensor Systems (ACM SenSys'03)*, 324-325, Los Angeles, USA.
6. Estrin, D., R. Govindan, J. S. Heidemann, and S. Kumar. 1999. "Next Century Challenges: Scalable Coordination in Sensor Networks." In *Proceedings of the ACM International Conference on Mobile Computing and Networking (MobiCom'99)*, 263-270, Seattle, Washington, USA, August 1999.
7. Hu, L., and D. Evans. 2003. "Secure Aggregation for Wireless Networks." In *Proceedings of theInternational Symposium on Applications and the Internet (SAINT'03) Workshops*, 384, Orlando, Florida, USA, January 2003, IEEE Computer Society.
8. Madden, S. M. J. Franklin, J. M. Hellerstein, and W. Hong. 2002. "TAG: A Tiny Aggregation Service for Ad-hoc Sensor Networks." *ACM SIGOPS Operating Systems Review* (Special Issue): 131-146.
9. Przydatek, B., D. Song, and A. Perrig. 2003. "SIA: Secure Information Aggregation in Sensor Networks." In *Proceedings of the 1st International Conference on Embedded Networked Systems (SenSys'03)*, 255-265, New York: ACM Press.
10. Shrivastava, N., C. Buragohain, D. Agrawal, and S. Suri. November 2004. "Medians and Beyond: New Aggregation Techniques for Sensor Networks." In *Proceedings of the 2nd International Conference on Embedded Networked Sensor Systems (ACM SenSys'04)*, 239-249, Baltimore, Maryland, USA.
11. Ye, F., L. H. Luo, and S. Lu. March 2004. "Statistical En-Route Filtering of Injected False Data in Sensor Networks." In *Proceedings of the 23rd IEEE Joint Annual Conference of Computer and Communication Societies (IEEE INFOCOM'04)*, vol 4, 2446-2457, Hong Kong, China.
12. Ganeriwal S., and M. Srivastava. 2004. "Reputation-Based Framework for High Integrity Sensor Networks." In *Proceedings of the 2nd ACM Workshop on Security on Ad Hoc and Sensor Networks (SASN'04)*, 66-77, Washington DC, USA.
13. Liang Z., and W. Shi. 2005. "Enforcing Cooperative Resource Sharing in Untrusted Peer-to-Peer Environment." *ACM Journal of Mobile Networks and Applications (MONET)*, 10 (6): 771-783.
14. Liang, Z., and W. Shi. February 2005. "Analysis of Ratings on Trust Inference in the Open Environment." Technical Report MIST-TR-2005-002, Department of Computer Science, Wayne State University, USA.
15. Liang Z., and W. Shi. January 2005. "PET: A Personalized Trust Model with Reputation and Risk Evaluation for P2P Resource Sharing." In *Proceedings of the 38th Annual Hawaii International Conference on System Sciences (HICSS)*, 201-202, Hilton Waikoloa Village Big Island, Hawaii.
16. Ren, K., T. Li, Z. Wan, F. Bao, R. H. Deng, and K. Kim. August 2004. "Highly Reliable Trust Establishment Scheme in Ad hoc Networks." *Computer Networks: The International Journal of Computer and Telecommunications Networking* 45: 687-699.
17. Tanachaiwiwat, S., P. Dave, R. Bhindwale, and A. Helmy. April 2004. "Location-Centric Isolation of Misbehavior and Trust Routing in Energy-Constrained Sensor Networks." In *Proceedings of IEEE International Conference on Performance, Computing, and Communications*, 463-469.
18. Yan, Z., P. Zhang, and T. Virtanen. October 2003. "Trust Evaluation Based Security Solution in Ad hoc Networks." In *Proceedings of the 7th Nordic Workshop on Secure IT Systems(NordSec'03)*, Gjovik, Norway.

19. Zhu, H., F. Bao, R. H. Deng, and K. Kim. September 2004. "Computing of Trust in Wireless Networks." In *Proceedings of 60$^{th}$ IEEE Vehicular Technology Conference,* Los Angeles, California, USA.
20. Anderson, R., and M. Kuhn. November 1996. "Tamper Resistance- A Cautionary Note." In *Proceedings of the 2$^{nd}$ USENIX Workshop on Electronic Commerce (WOEC'96)*, 1-11, Oakland, California, USA.
21. Anderson, R., and M. Kuhn.1997. "Low Cost Attacks on Tamper Resistant Devices." In *Proceedings of the 5th International Workshop on Security Protocols (IWSP), Lecture Notes in Computer Science(LNCS)*, Vol 1361, 125-136.
22. Hartung, C., J. Balasalle, and R. Han. 2004. "Node Compromise in Sensor Networks: The Need for Secure Systems." Technical Report CU-CS-988-04, Department of Computer Science, University of Colorado at Boulder, Colorado, USA.
23. Hu L., and D. Evans. February 2004. "Using Directional Antennas to Prevent Wormhole Attacks." In *Proceedings of the 11$^{th}$ Annual Network and Distributed System Security Symposium(NDSS'04)*, 131-41, San Diego, California, USA.
24. Komerling O., and M. G. Kuhn. 1999. "Design Principles for Tamper-Resistant Smart Card Processors." In *Proceedings of USENIX Workshop on Smartcard Technology*, 9-20, Chicago, Illinois, USA.
25. Sastry, N., U. Shankar, and D. Wagner. September 2003. "Secure Verification of Location Claims." In *Proceedings of the 2$^{nd}$ ACM Workshop on Wireless Security,* 1-10, Sandiego, California, USA.
26. Seshadri, A., A. Perrig, L. Van Doorn, and P. Khosla. May 2004. "SWATT: Software-Based Attestation for Embedded Devices." In *Proceedings of the IEEE Symposium on Security and Privacy,* 272-282, Oakland, California, USA.
27. Wang, X., W. Gu, S. Chellappan, K.Schoseck, and D. Xuan. May 2005. "Lifetime Optimization of Sensor Networks under Physical Attacks." In *Proceedings of IEEE International Conference on Communications(ICC)*, Vo 5, 3295-301.
28. Wang, X., W. Gu, S. Chellappan, Dong Xuan, and Ten H. Laii. February 2005. "Search-Based Physical Attacks in Sensor Networks: Modeling and Defense." Technical Report, Department of Computer Science and Engineering, Ohio State University.
29. Wood A. D., and J.A. Stankovic. 2002. "Denial of Service in Sensor Networks." *IEEE Computer,* 35 (10), 54-62.
30. Carman, D. W., P. S. Krus, and B. J. Matt. 2000. "Constraints and Approaches for Distributed Sensor Network Security." Technical Report 00-010, NAI Labs, Network Associates Inc., Glenwood, MD, USA.
31. Hill, J., R. Szewczyk, A. Woo, S. Hollar, D. E. Culler, and K. Pister. 2000. "System Architecture Directions for Networked Sensors." In *Proceedings of the 9$^{th}$ International Conference on Architectural Support for Programming Languages and Operating Systems,* 93-104, New York : ACM Press.
32. Slijepcevic, S., M. Potkonjak, V. Tsiatsis, S. Zimbeck, and M. B. Srivastava. June 2002. "On Communication Security in Wireless Ad-hoc Sensor Networks." In *Proceedings of 11$^{th}$ IEEE International Workshop on Enabling Technologies: Infrastruture for Collaborative Enterprises (WETICE'02),* 139-144, Pittsburg, Pennsylvania, USA.
33. Yuan L., and G. Qu. July 2002. "Design Space Exploration for Energy-Efficient Secure Sensor Networks." In *Proceedings of IEEE International Conference on Application-Specific Systems, Architectures, and Processors(ASAP'02)*, 88-100, San Jose, California, USA.
34. URL: http://www.willow.co.uk/html/telosb_mote_platform.html. 2010. (Accessed on July 11, 2012).
35. Perrig, A., R. Szewczyk, V. Wen, D. E. Culler, and J. D. Tygar. 2002. "SPINS: Security Protocols for Sensor Networks." *Wireless Networks*, 8 (5): 521-534.
36. J.A. Stankovic, J. A. , T. Abdelzaher, C. Lu, L. Sha, and J. Hou. July 2003. "Real-Time Communication and Coordination in Embedded Sensor Networks." In *Proceedings of the IEEE,* 91(7): 1000-1022.
37. Eschenauer L., and V. D. Gligor. November 2002. "A Key-Management Scheme for Distributed Sensor Networks." In *Proceedings of the 9$^{th}$ ACM Conference on Computer and Communications Security (CCS'02),* 41-47, Washington DC, USA.
38. Chan, H., A. Perrig, and D. Song. May 2003. Random Key Pre-Distribution Schemes for Sensor Networks." In *Proceedings of the IEEE Symposium on Security and Privacy (S&P'03)*, 197, Berkeley, California, USA.
39. Hwang J., and Y. Kim. 2004. "Revisiting Random Key Pre-Distribution Schemes for Wireless Sensor Networks." In *Proceedings of the 2$^{nd}$ ACM Workshop on Security of Ad Hoc and Sensor Networks (SASN'04),* 43-52, ACM Press, New York.
40. Liu, D., P. Ning, and R. Li. 2005. "Establishing Pairwise Keys in Distributed Sensor Networks." *ACM Transactions on Information Systems Security*, 8 (1): 41-77.
41. Capkun S., and J.-P. Hubaux. 2006. "Secure Positioning in Wireless Networks." *IEEE Journal on Selected Areas in Communications,* 24 (2): 221-232.

42. Lazos L., and R. Poovendran. 2005. "SERLOC: Robust Localization for Wireless Sensor Networks." *ACM Transactions on Sensor Networks,* 1 (1): 73-100.
43. Ganeriwal, S., S. Capkun, C.- C. Han, and M. B. Srivastava. 2005. "Secure Time Synchronization Service for Sensor Networks." In *Proceedings of the 4$^{th}$ ACM Workshop on Wireless Security,* 97-106, ACM Press, New York.
44. Shi E., and A. Perrig. December 2004. "Designing Secure Sensor Networks." *Wireless Communication Magazine*, 11 (6): 38-43.
45. Wang, X., W. Gu, K. Schosek, S. Chellappan, and D. Xuan. July 2004. "Sensor Network Configuration under Physical Attacks." Technical Report (OSU-CISRC-7/04-TR45), Department of Computer Science and Engineering, Ohio State University, Ohio, USA.
46. Karlof C., and D. Wagner. May 2003. "Secure Routing in Wireless Sensor Networks: Attacks and Countermeasures." In *Proceedings of the 1st IEEE International Workshop on Sensor Network Protocols and Applications,* 113-127, Anchorage, Alaska, USA.
47. Newsome, J., E. Shi, D. Song, and A. Perrig. 2004. "The Sybil Attack in Sensor Networks: Analysis and Defenses." In *Proceedings of the 3$^{rd}$ International Symposium on Information Processing in Sensor Networks*, 1259-1268.
48. Douceur, J. March 2002. "The Sybil Attack." In *Proceedings of the 1$^{st}$ International Workshop on Peer-to-Peer Systems (IPTPS'02),* 251-260, Cambridge Massachusetts, USA, Springer LNCS, Vol. 2429.
49. Awerbuch, B., D. Holmer, C. Nita-Rotaru, and H. Rubens. September 2002. "An On-Demand Secure Routing Protocol Resilient to Byzantine Failures." In *Proceedings of the 1$^{st}$ ACM Workshop on Wireless Security (WiSe'02)*, 21-30, Atlanta, Georgia, USA.
50. Sen, J. December 2010. "Routing Security Issues in Wireless Sensor Networks: Attacks and Defense." In *Sustainable Wireless Sensor Networks*, edited by Y. K. Tan, Chapter 12, 279 – 309, Croatia, INTECH Publishers.
51. Parno, B., A. Perrig, and V. Gligor. May 2005. "Distributed Detection of Node Replication Attacks in Sensor Networks." In *Proceedings of the IEEE Symposium on Security and Privacy (S&P'05),* 49-63, Oakland, California, USA.
52. Gruteser, M., G. Schelle, A. Jain, R. Han, and D. Grunwald. May 2003. "Privacy-Aware Location Sensor Networks." In *Proceedings of the 9$^{th}$ USENIX Workshop on Hot Topics in Operating Systems (HotOS IX),* Vol 9, 28, Lihue, Hawaii, USA.
53. Ozturk, C., Y. Zhang, and W. Trappe. October 2004. "Source-Location Privacy in Energy-Constrained Sensor Network Routing." In *Proceedings of the 2$^{nd}$ ACM Workshop on Security of Ad Hoc and Sensor Networks(SASN'04),* 88-93, Washington DC, USA.
54. Chan H., and A. Perrig. 2003. "Security and Privacy in Sensor Networks." *IEEE Computer Magazine*, 36 (10): 103-105.
55. Deng, J., R. Han, and S. Mishra. 2004. "Countermeasures against Traffic Analysis in Wireless Sensor Networks." Technical Report CU-CS-987-04, University of Colorado at Boulder, 2004.
56. Perrig, A., J. Stankovic, and D. Wagner. 2004. "Security in Wireless Sensor Networks." *Communications of ACM* , 47 (6): 53-57.
57. Malan, D. J., M. Welsh, and M.D. Smith. October 2004. "A Public-Key Infrastructure for Key Distribution in TinyOS based on Elliptic Curve Cryptography." In *Proceedings of the 1$^{st}$ IEEE International Conference on Sensor and Ad Hoc Communications and Networks,* Santa Clara, California, October, 2004.
58. Rivest, R. L., A. Shamir, and L. Adleman. 1983. "A Method for Obtaining Digital Signatures and Public-Key Cryptosystems." *Communications of the ACM,* 26 (1): 96-99.
59. Brown, M., D. Cheung, D. Hankerson, J.L. Hernandez, M. Kirkup, and A. Menezes. August 2000. "PGP in Constrained Wireless Devices." In *Proceedings of the 9$^{th}$ USENIX Security Symposium (SSYM'00),* Vol 9, 19.
60. Gura, N., A. Patel, A. Wander, H. Eberle, and S. Shantz. August 2004. "Comparing Elliptic Curve Cryptography and RSA on 8-bit CPUs." In *Proceedings of the 6$^{th}$ International Workshop on Cryptographic Hardware and Embedded Systems (CHES '04),* 119-132, Cambridge, Massachusetts, USA., Springer LNCS Vol. 3156.
61. Gaubatz, G., J. P. Kaps, and B. Sunar. August 2004. "Public Key Cryptography in Sensor Networks-Revisited." in *Proceedings of the 1$^{st}$ European Workshop on Security in Ad-Hoc and Sensor Networks (ESAS'04),* 2-18, Heidelberg, Germany, Springer LNCS, Vol. 3313.
62. Wander, A. S., N. Gura, H. Eberle, V. Gupta, and S.C. Shantz. March 2005. "Energy Analysis of Public-Key Cryptography for Wireless Sensor Networks." In *Proceedings of the 3$^{rd}$ IEEE International Conference on Pervasive Computing and Communication (PERCOM'05),* 324-328, Kauai, Island, Hawaii, USA.

63. Rabin M. O. 1979. "Digitalized Signatures and Public-Key Functions as Intractable as Factorization", Cambridge, MA, Technical Report.
64. Hoffstein, J., J. Pipher, and J. H. Silverman. June 1998. "NTRU: A Ring-Based Public Key Cryptosystem." In *Proceedings of the 3$^{rd}$ International Symposium on Algorithmic Number Theory(ANTS'98),* 267-288, Portland, Oregon, USA. Springer LNCS, Vol 1423.
65. Miller, V. S. August 1986. "Use of Elliptic Curves in Cryptography." In *Proceedings of the Advances in Cryptology- CRYPTO'85*, 417-426, Santa Barbara, Califiornia, USA, Springer LNCS, Vol. 218.
66. Kobiltz, N. 1987. "Elliptic Curve Cryptosystems." *Mathematics of Computation*, 48: 203-209.
67. Elliptic Curve Cryptography, SECG Std. SEC1, 2000. http://www.secg.org/collateral/sec1.pdf. Accessed on July 11, 2012.
68. Kaliski, B. May 2003. TWIRL and RSA Key Size, RSA Laboratories, Technical Note.
69. Recommended Elliptic Curve Domain Parameters, SECG Std. SEC 2, 2000. http://www.secg.org/collateral/sec2_final.pdf. Accessed on July 11, 2012..
70. Hankerson, D., A.Menezes, and S. Vanstone. 2004. *Guide to Elliptic Curve Cryptography*, New York, Springer-Verlag.
71. Freier, A., P. Karlton, and P. Kocher. *The SSL Protocol*, version 3.0.. http://www.mozilla.org/projects/security/pki/nss/ssl/draft302.text. Accessed on July 11, 2012.
72. Watro, R., D. Kong, S. Cuti, C. Gardiner, C. Lynn, and P. Kruus. 2004. "TinyPK: Securing Sensor Networks with Public Key Technology." In *Proceedings of the 2$^{nd}$ ACM Workshop on Security of Ad Hoc and Sensor Networks (SASN'04)*, 59-64, New York: ACM Press.
73. Liu A., and P. Ning. April 2008. "TinyECC: A Configurable Library for Elliptic Curve Cryptography in Wireless Sensor Networks." In *Proceedings of the 7$^{th}$ International Conference on Information Processing in Sensor Networks (IPSN'08)*, SPOTS Track, 245-256, St Louis, Missouri, USA. http://discovery.csc.ncsu.edu/software/TinyECC/. Accessed on July 11, 2012.
74. Karlof, C., N. Sastry, and D. Wagner. November 2004. "TinySec: A Link Layer Security Architecture for Wireless Sensor Networks." In *Proceedings of the 2$^{nd}$ ACM Conference on Embedded Networked Sensor Systems (SensSys'04),* 162-175, Baltimore, MD, USA.
75. U.S. National Institute of Standards and Technology (NIST). June 1998. SKIPJACK and KEA Algorithm Specifications. Federal Information Processing Standards Publications 185 (FIPS PUB 185).
76. Rivest, R. L. 1995. "The RC5 Encryption Algorithm." In *Proceedings of the International Workshop on Fast Software Encryption*, 86-96, Springer LNCS Vol 1008.
77. Eastlake D., and P. Jones. September 2001. "U.S. Secure Hash Algorithm 1 (SHA1)"*,* RFC 3174 (Informational).
78. Daemen J., and V. Rijmen. August 1998. "AES Proposal: Rijndael." In *Proceedings of 1$^{st}$ AES Candidate Conference (AES1),* Ventura, California, USA.
79. Menezes, A. J., S. A. Vanstone, and P.C.V. Oorschot. 1996. *Handbook of Applied Cryptography*, Boca Raton, Florida, USA, CRC Press.
80. R.L. Rivest. April 1992. *The MD5 Message-Digest Algorithm*, RFC 1321.
81. Ganesan, P., R. Venugopalan, P. Peddabachagari, A. Dean, F. Mueller, and M. Sichitiu. 2003. "Analyzing and Modeling Encryption Overhead for Sensor Network Nodes." In *Proceedings of the 2$^{nd}$ ACM International Conference on Wireless Sensor Networks and Applications,* 151-159, New York: ACM Press.
82. Law, Y. W., J. M. Doumen, and P. H. Hartel. October 2004. "Benchmarking Block Ciphers for Wireless Sensor Networks (Extended Abstract)." In *Proceedings of the 1$^{st}$ IEEE International Conference of Mobile Ad-hoc and Sensor Systems,* 447- 456,IEEE Computer Society Press.
83. Wheeler D. J., and R.M. Needham. December 1994. "TEA: A Tiny Encryption Algorithm." In *Proceedings of Fast Software Encryption: 2$^{nd}$ International Workshop,* 363-366, Leuven, Belgium. Springer LNCS, Vol. 1008.
84. Rivest, R. L., M. J. B. Robshaw, R. Sidney, and Y.L. Yi. "The RC6 Block Cipher". ftp://ftp.rsasecurity.com/pub/rsalabs/rc6/rc6v11.pdf. Accessed on July 11, 2012.
85. Matsui, M. January 2007. "New Block Encryption Algorithm MISTY." In *Proceedings of the 4$^{th}$ International Workshop of Fast Software Encryption (FSE'97),* 54-58, Haifa, Israel. Springer LNCS Vol 1267.
86. 3GPP Specification Detail 2011: 3G Security: Specification of the 3GPP Confidentiality and Integrity Algorithms: Document 2: KASUMI Specification. http://www.3gpp.org/ftp/Specs/html-info/35202.html.
87. Aoki, K., T. Ichikawa, M. Matsui, S. Moriai, J. Nakajima, and T. Tokita. 2001. Specification of Camellia- A 128-bit Block Cipher, Specification (Version 2.0). Nippon Telegraph and Telephone Corporation and Mitsubishi Electric Corporation.

88. Di Pietro, R., L.V. Mancini, Y.W. Law, S. Etalle, and P. Havinga. October 2003. "LKHW: A Directed Diffusion-Based Secure Multi-Cast Scheme for Wireless Sensor Networks." In *Proceedings of the 32$^{nd}$ International Conference on Parallel Processing Workshops (ICPPW'03)*, 397-406, Kaohsiung, Taiwan, IEEE Computer Society.
89. Zhu, S., S. Setia, and S. Jajodia. 2003. "LEAP: Efficient Security Mechanism for Large –Scale Distributed Sensor Networks." In *Proceedings of the 10$^{th}$ ACM Conference on Computer and Communications Security*, 62-72, New York: ACM Press.
90. Lai, B., S. Kim, and I. Verbauwhede. 2002. "Scalable Session Key Construction Protocols for Wireless Sensor Networks." In *Proceedings of the IEEE Workshop on Large Scale Real Time and Embedded Systems (LATES'02)*, 1-6, Austin, Texas, USA.
91. Cametepe, S. A., and B. Yener. 2007. "Combinatorial Design of Key Distribution Mechanisms for Wireless Sensor Networks." *IEEE/ACM Transactions on Networksing (TON)*, 15 (2): 346 – 358.
92. Lee J., and D.R. Stinson. August 2004. "Deterministic Key Pre-Distribution Schemes for Distributed Sensor Networks." In Proceedings of the 11$^{th}$ International Workshop on Selected Areas in Crypography (SAC'04), 294-307, Waterloo, Canada, *Selected Areas in Cryptography,* pp. 294-307. Springer LNCS Vol. 3357.
93. Lee J., and D.R. Stinson. March 2005. "A Combinatorial Approach to Key Pre-Distribution for Distributed Sensor Networks." In *Proceedings of the IEEE Wireless Communications and Networking Conference (WCNC'05),* Vol. 2, 1200-1205, New Orleans, Los Angeles, USA.
94. Chan H., and A. Perrig. March 2005. "PIKE: Peer Intermediaries for Key Establishment in Sensor Networks." In *Proceedings of the 25$^{th}$ IEEE Annual International Conference on Computer and Communications (INFOCOM'05),* 525-535, Miami, Florida, USA.
95. Huang, Q., J. Cukier, H. Kobayashi, B. Liu, and J. Zhang. 2003. "Fast Authenticated Key Establishment Protocols for Self-Organizing Sensor Networks." In *Proceedings of the 2$^{nd}$ ACM International Conference on Wireless Sensor Networks and Applications (WSNA'03),* 141-150, San Diego, California, USA.
96. Zhou Y., and Y. Fang. April 2006. "A Scalable Key Agreement Scheme for Large Scale Networks." In *Proceedings of IEEE International Conference on Networking, Sensing and Control (ICNSC'06),* 631-636, Fort Lauderdale, Florida, USA.
97. Du, W., J. Deng, Y. S. Han, and P. K. Varshney. 2003. "A Pair-Wise Key Pre-Distribution Scheme for Wireless Sensor Networks." In *Proceedings of the 10$^{th}$ ACM Conference on Computer and Communications Security*, 42-51, New York: ACM Press.
98. Di. Pietro R., L. V. Mancini, and A. Mei. 2003. "Random Key-Assignment for Secure Wireless Sensor Networks." In *Proceedings of the 1$^{st}$ ACM Workshop on Security of Ad hoc and Sensor Networks,* 62-71, New York: ACM Press.
99. Du, W., J. Deng, Y.S. Han, S. Chen, and P. K. Varshney. 2004. "A Key Management Scheme for Wireless Sensor Networks Using Deployment Knowledge." In *Proceedings of IEEE INFOCOM,* 586-597, Hong Kong, China.
100. Hwang, D. D., B. Lai, and I. Verbauwhede. July 2004. "Energy-memory-security tradeoffs in distributed sensor networks, in *Proceedings of the 3$^{rd}$ International Conference on Ad-hoc Networks and Wireless (ADHOC-NOW),* 70-81. Springer LNCS, Vol. 3158.
101. Blundo, C., A. D. Santis, A. Herzberg, S. Kutten, U. Vaccaro, and M. Yung. 1998. "Perfectly-Secure Key Distribution for Dynamic Conferences." *Information and Computation*, 146 (1): 1 – 23.
102. Liu D., and P. Ning. October 2003. "Location-Based Pair-Wise Key Establishments for Static Sensor Networks." In *Proceedings of the ACM Workshop on Security in Ad hoc and Sensor Networks,* 72-82.
103. Zhang, Y., J. Zheng, and H. Hu. 2008. *Security in Wireless Sensor Networks,* CRC Press, Taylor & Francis Group, Boca Raton, Florida, USA.
104. Chan, H., V. Gligor, A. Perrig, and G. Muralidharan. July 2005. "On the Distribution and Revocation of Cryptographic Keys in Sensor Networks." *IEEE Transactions on Dependable and Secure Computing,* 2 (3): 233-247.
105. Sen, J., M. G. Chandra, P. Balamuralidhar, S. G. Harihara, H. Reddy. May 2007. "A Distributed Protocol for Detection of Packet Dropping Attack in Mobile Ad hoc Networks." In *Proceedings of the International Conference on Telecommunications and Malaysian International Conference on Communications (ICT-MICC'07)*, Penang, Malysia.
106. Hu, Y., A. Perrig, and D.B. Jonson. March 2003. "Packet Leashes: A Defense against Worm-Hole Attacks." In *Proceedings of the 22$^{nd}$ Annual Joint Conference of the IEEE Computer and Communications Socities (INFOCOM'03),* Vol 3, 1976- 1986, San Francisco, California, USA.
107. Wang W., and B. Bhargava. 2004. "Visualization of Wormholes in Sensor Networks." In *Proceedings of the 2004 ACM Workshop on Wireless Security,* 51-60, New York: ACM Press.

108. Sen, J., M. G. Chandra, P. Balamuralidhar, S. G. Harihara, H. Reddy. December 2007. "A Mechanism for Detection of Gray Hole Attack in Mobile Ad hoc Networks." In *Proceedings of the 6th International Conference on Information, Communications and Signal Processing (ICICS'07)*, Singapore.
109. Aura, T., P. Nikander, and J. Leiwo. April 2000. "DOS-Resistant Authentication with Client Puzzles." In *Proceedings of the 8th International Workshop on Security Protocols*, 170-177, Cambridge, UK. Springer LNCS, Vol. 2133.
110. Rafaeli S., and D. Hutchison. 2003. "A Survey of Key Management for Secure Group Communications." *ACM Computing Survey,* 35 (3): 309-329.
111. Lazos L., and R. Poovendran. April 2003. "Energy-Aware Secure Multi-Cast Communication in Ad-hoc Networks Using Geographic Location Information." In *Proceedings of the IEEE International Conference on Acoustics Speech and Signal Processing(ICASSP),* Vol 4, 201-204, Hong Kong, China.
112. Lazos L., and R. Poovendran. September 2002. "Secure Broadcast in Energy-Aware Wireless Sensor Networks." In *Proceedings of the IEEE International Symposium on Advances in Wireless Communications (ISWC'02),* 1, Invited Paper, Victoria, British Columbia, Canada. .
113. Intanagonwiwat, C., R. Govindan, and D. Estrin. August 2000. "Directed Diffusion: A Scalable and Robust Communication Paradigm for Sensor Networks." In *Proceedings of the 6th ACM Annual International Conference on Mobile Computing and Networking (MobiCom'00)*, 56-67, Boston, Massachusetts, USA. .
114. Kaya, T., G. Lin, G. Noubir, and A. Yilmaz. October 2003. "Secure Multicast Groups on Ad hoc Networks." In *Proceedings of the 1st ACM Workshop on Security of Ad Hoc and Sensor Networks (SASN'03),* 94-102, Fairfax, Virginia, USA.
115. Al-Karaki J. N., and A. E. Kamal. December 2004. "Routing Techniques in Wireless Sensor Networks: A Survey." *IEEE Wireless Communications,* 11 (6): 6-28.
116. Du, W., R. Wang, and P. Ning. 2005. "An Efficient Scheme for Authenticating Public Keys in Sensor Networks." In *Proceedings of the 6th ACM International Symposium on Mobile Ad hoc Networking and Computing,* 58-67, New York: ACM Press.
117. Liu D., and P. Ning. February 2003. "Efficient Distribution of Key Chain Commitments for Broadcast Authentication in Distributed Sensor Networks." In *Proceedings of the 10th Annual Network and Distributed System Security Symposium*, 263-276, San Diego, California, USA.
118. Liu D., and P. Ning. 2004. "Multilevel µTESLA: Broadcast Authentication for Distributed Sensor Networks." *ACM Transactions on Embedded Computing Systems (TECS),* 3 (4): 800-836.
119. Liu, D., P. Ning, S. Zhu, and S. Jajodia. July 2005. "Practical Broadcast Authentication in Sensor Networks." In *Proceedings of the 2nd Annual International Conference on Mobile and Ubiquitous Systems: Networking and Services (MobiQuitous'05)*, 118-129, San Diego, California, USA.
120. Public-Key Infrastructure (X.509) (pkix). http://www.ietf.org/html.charters/pkix-charter.html. Accessed On July 11, 2012.
121. Sen J., and A. Ukil. 2010. "A Secure Routing Protocol for Wireless Sensor Networks." In *Proceedings of the International Conference on Computational Science and its Application (ICCSA'10)*, Fukuoka, Japan, Vol. 3, 277 – 290, Springer LNCS 6018, Heidelberg, Germany.
122. Braginsky, D., and D. Estrin. 2002. "Rumor Routing Algorithm for Sensor Networks." In *Proceedings of the 1st ACM International Workshop on Wireless Sensor Networks and Applications,* 22-31, New York: ACM Press.
123. Gruteser M., and D. Grunwald. May 2003. "Anonymous Usage of Location-Based Services through Spatial and Temporal Cloaking." In *Proceedings of the 1st International Conference on Mobile Systems, Applications, and Services (MobSys'03)*,31-42, San Francisco, California, USA.
124. Beresford A. R., and F. Stajano. 2003. "Location Privacy in Pervasive Computing." *IEEE Pervasive Computing*, 2 (1): 46-55.
125. Sen, J. December 2010. "An Efficient and User Privacy-Preserving Routing Protocol for Wireless Mesh Networks." *International Journal on Scalable Computing: Practice and Experience, Special Issue on Network and Distributed Systems*, 11 (4): 345-358.
126. Rivest, R., A. Shamir, and Y. Tauman. December 2001. "How to Leak a Secret." In *Proceedings of the 7th International Conference on the Theory and Applications of Cryptology and Information Security (ASIACRYPT'01)*, 552-565, Gold Coast, Australia. Springer LNCS, Vol. 2249.
127. Gruteser M., and D. Grunwald. March 2003. "A Methodological Assessment of Location Privacy Risks in Wireless Hotspot Networks." In *Proceedings of the 1st International Conference on Security in Pervasive Computing (SPC'03),* 10-24, Boppard, Germany. Springer LNCS, Vol. 2802.
128. Priyantha, N. B., A. Chakraborty, and H. Balakrishnan. August 2000. "The Cricket-Location Support System." In *Proceedings of the 6th Annual ACM International Conference on Mobile Computing and Networking (MobiCom'00),* 32-43, Boston, Massachusetts, USA.

129. Smailagic, A., D. P. Siewiorek, J. Anhalt, and Y. Wang, D. Kogan. 2001. "Location Sensing and Privacy in a Context Aware Computing Environment." *IEEE Wireless Communications,* 9: 10-17.
130. Molnar D., and D. Wagner. October 2004. "Privacy and Security in Library RFID: Issues, Practices, and Architectures." In *Proceedings of the 11th ACM Conference on Computer and Communications Security (CCS'04)*, 204- 219, Washington DC, USA.
131. Duri, S., M. Gruteser, X. Liu, P. Moskowitz, R. Perez, M. Singh, and J. Tang. September 2000. "Framework for Security and Privacy in Automotive Telematics." In *Proceedings of the 2nd ACM International Workshop on Mobile Commerce (WMC'02)*, 25-32, Atlanta, Georgia, USA.
132. Snekkenes, E. October 2001. "Concepts for Personal Location Privacy Policies." In *Proceedings of the 3rd ACM Conference on Electronic Commerce (ACM-EC'01),* 48-57, Tampa, Florida, USA.
133. Myles, G., A. Friday, and N. Davies. 2003. "Preserving Privacy in Environments with Location-Based Applications." *IEEE Pervasive Computing,* 2 (1): 56-64.
134. Hengartner U., and P. Steenkiste. March 2003. "Protecting Access to People Location Information." In *Proceedings of the 1st International Conference on Security in Pervasive Computing (SPC'03)*, 222-231, Boppard, Germany. Springer LNCS, Vol 2802.
135. Xi, Y., L. Schwiebert, and W. Shi. April 2006. "Preserving Privacy in Monitoring-Based Wireless Sensor Networks." In *Proceedings of the 2nd International Workshop on Security in Systems and Networks (SSN'06),* Rhode Island, Greece.
136. Sato, I., Y. Okazaki, and S. Goto. 2002. "An Improved Intrusion Detection Method Based on Process Profiling.", *IPSJ Journal,* 43 (11): 3316-3326.
137. Marti, S., T. J. Giuli, K. Lai, and M. Baker. 2000. "Mitigating Routing Misbehavior in Mobile Ad hoc Networks." In *Proceedings of the 6th Annual International Conference on Mobile Computing and Networking,* 255-265, New York: ACM Press.
138. Zhang, Y., W. Lee, and Y.-A. Huang. 2003. "Intrusion Detection Techniques for Mobile Wireless Networks." *Wireless Networks.* 9 (5): 545-556.
139. Huang, Y., W. Fan, W. Lee, and P. S. Yu. May 2003. "Cross-Feature Analysis for Detecting Ad-hoc Routing Anomalies." In *Proceedings of the 23rd International Conference on Distributed Computing Systems (ICDCS'03),* 478, Providence, RI.
140. Huang Y., and W. Lee. September 2004. "Attack Analysis and Detection for Ad hoc Routing Protocols." In *Proceedings of the 7th International Symposium on Recent Advances in Intrusion Detection,* 125-145, Sophia Antipolis, France.
141. Sen, J., I. Sengupta. December 2005. "Autonomous Agent-Based Distributed Fault-Tolerant Intrusion Detection System." In *Proceedings of the 2nd International Conference on Distributed Computing and Internet Technology (ICDCIT '05),* 125-131, Bhubaneswar, India. Springer LNCS Vol 3186.
142. Brutch, P., and C. Ko. January 2003. "Challenges in Intrusion Detection for Wireless Ad-hoc Networks." In *Proceedings of the Symposium on Applications and the Internet Workshops (SAINT'03 Workshops)*, 368, Orlando, Florida, USA.
143. Zhu, S., S. Setia, S. Jajodia, and P. Ning. May 2004. "An Interleaved Hop-by-Hop Authentication Scheme for Filtering of Injected False Data in Sensor Networks." In *Proceedings of IEEE Symposium on Security and Privacy (S&P'04),* 259-271, Oakland, California, USA.
144. Wang, G., W. Zhang, C. Cao, and T.L. Porta. October 2003. "On Supporting Distributed Collaboration in Sensor Networks." In *Proceedings of the IEEE Millitary Communications Conference (MILCOM'03),* Vol 2, 752-757.
145. Albers, P., and O. Camp. April 2002. "Security in Ad hoc Networks: A General Intrusion Detection Architecture Enhancing Trust-Based Approaches." In *Proceedings of the 1st International Workshop on Wireless Information Systems (WIS'02),* 1-12, Ciudad Real, Spain.
146. Merkle, R. C. April 1980. "Protocols for Public Key Cryptosystems." In *Proceedings of the IEEE Symposium on Research in Security and Privacy (S&P'80),* 122-134, Oakland, California, USA.
147. Deng, J., R. Han, and S. Mishra. 2003. "Security Support for In-Network Processing in Wireless Sensor Networks." In *Proceedings of the 1st ACM Workshop on Security of Ad hoc and Sensor Networks,* 83-93, New York: ACM Press.
148. H. Cam, D. Muthuavinashiappan, and P. Nair, "ESPDA: Energy-efficient and secure pattern-based data aggregation for wireless sensor networks", in *Proceedings of IEEE Sensors,* pp. 732-736, Toronto, Canada, October 2003.
149. Cam, H., D. Muthuavinashiappan, and P. Nair. October 2003. "Energy-Efficient Security Protocol for Wireless Sensor Networks." In *Proceedings of IEEE VTC Conference,* 2981-2984, Orlando, Florida, USA.

150. Cam, H., S. Ozdemir, H. O. Sanli, and P. Nair. May 2004. "Secure Differential Data Aggregation for Wireless Sensor Networks." In: *Sensor Network Operations*, S. Phoha, T. F. La Porta, and C. Griffin (Eds.), Willy-IEEE Press.
151. Du, W., J. Deng, Y. S. Han, and P. K. Varshney. December 2003. "A Witness-Based Approach for Data Fusion Assurance in Wireless Sensor Networks." In *Proceedings of IEEE Global Telecommunications Conference (GlobeCom'03),* 1435-1439, San Francisco, USA.
152. Wagner, D. 2004. "Resilient Aggregation in Sensor Networks." In *Proceedings of the 2$^{nd}$ ACM Workshop on Security of Ad Hoc and Sensor Networks (SASN'04),* 78-87, New York, NY, USA: ACM Press.
153. Acharya, M., J. Girao, and D. Westhoff. 2005. "Secure Comparison of Encrypted Data in Wireless Sensor Networks." In *Proceedings of the 3$^{rd}$ International Symposium on Modeling and Optimization in Mobile, Ad Hoc and Wireless Networks (WIOPT)*, 47-53, Washington, DC, USA.
154. Catelluccia, C., A. C- F. Chan, E. Mykletun, and G. Tsudik. May 2009. "Efficient and Provably Secure Aggregation of Encrypted Data in Wireless Sensor Networks." *ACM Transactions on Sensor Networks (TOSN)*, 5 (3), Article No: 20.
155. Girao, J., D. Westhoff, M. Schneider. May 2005. "CDA: Concealed Data Aggregation for Reverse Multicast Traffic in Wireless Sensor Networks." In *Proceedings of IEEE International Conference on Communications (ICC'05)*, 5: 3044-3049, Seoul, Korea.
156. He, W., X. Liu, H. Ngyyen, K. Nahrstedt, and T. Abdelzaher. May 2007. " PDA: Privacy-Preserving Data Aggregation in Wireless Sensor Networks." In *Proceedings of the 26$^{th}$ IEEE International Conference on Computer Communications (INFOCOM)'07)*, 2045 – 2053, Anchorage, Alaska, USA.
157. Westhoff, D., J. Girao, and M. Acharya. 2006. "Concealed Data Aggregation for Reverse Multicast Traffic in Sensor Networks: Encryption, Key Distribution, and Routing Adaptation." *IEEE Transactions on Mobile Computing,* 5 (10): 1417-1431.
158. Armknecht, F., D. Westhoff, J. Girao, and A. Hessler. 2008. "A Lifetime-Optimized End-to-End Encryption Scheme for Sensor Networks Allowing In-Network Processing." *Computer Communications*, 31 (4): 734-749.
159. Rivest, R. L., L. Adleman, and M. L. Dertouzos. 1978. "On Data Banks and Privacy Homomorphisms." In *Foundations of Secure Computation*, Workshop, Georgia Institute of Technology, Atlanta, 1977, 169-179, New York: Academic Press.
160. Domingo-Ferrer, J. September 2002. "A Provably Secure Additive and Multiplicative Privacy Homomorphism." In *Proceedings of the 5$^{th}$ International Conference on Information Security(ISC'02)*, 471-483, Sao Paulo, Brazil. Springer LNCS, Vol 2433.
161. Wagner, D. October 2003. "Cryptanalysis of an Algebraic Privacy Homomorphism." In *Proceedings of the 6$^{th}$ Information Security Conference* 234-239, Bristol, UK. Springer LNCS, Vol 2851.
162. Peter, S., D. Westhoff, and C. Castelluccia. 2010. "A Survey on the Encryption of Converge-Cast Traffic with In-Network Processing." *IEEE Transactions on Dependable and Secure Computing*, 7 (1): 20-34.
163. Deng, J., R. Han, and S. Mishra. July 2005. "Security, Privacy, and Fault-Tolerance in Wireless Sensor Networks." In: *Wireless Sensor Networks: A Systems Perspective*, N. Bulusu and S. Jha (Eds.), Artech House.
164. Sen, J. 2010. "Reputation- and Trust-Based Systems for Wireless Self-Organizing Networks.",In: *Security of Self-Organizing Networks: AMNET, WSN, WMN, VANET*, edited by Al-Sakib Khan Pathan, Aurbach Publications, Book Chapter No. 5, 91- 124, CRC Press, Taylor & Francis Group, USA.
165. Pirzada A., and C. McDonald. 2004. "Establishing Trust in Pure Ad hoc Networks." In *Proceedings of the 27$^{th}$ Australian Conference on Computer Science,* 47-54, Dunedin, New Zealand.
166. Oram, A. March 2001. *Peer-to-Peer: Harnessing the Power of Disruptive Technologies*. O'Reilly & Associates.
167. Sen, J. 2012. "Secure and Privacy-Aware Searching in Peer-to-Peer Networks." In *Proceedings of the 6$^{th}$ International Workshop on Data Privacy Management (DPM 2011)*, Leuven, Belgium, September 2011, edited by J. Garcia-Alfaro et al., Springer LNCS, Vol. 7122, 72-89, Heidelberg, Germany.
168. Sen, J. January 2011. "A Robust and Secure Aggregation Protocol for Wireless Sensor Networks." In *Proceedings of the 6$^{th}$ IEEE International Symposium on Electronic Design, Test and Applications (DELTA 2011)*, 222-227, Queenstown, New Zealand.
169. Sen, J. October 2010. "A Distributed Trust Management Framework for Detecting Malicious Packet Dropping Nodes in a Mobile Ad hoc Network." *International Journal of Network Security and its Applications (IJNSA)*, 2 (4): 92-104.
170. Sen, J. July 2010. "A Distributed Trust and Reputatation Framework for Mobile Ad hoc Networks." In *Proceedings of the 1$^{st}$ International Conference on Network Security and its Applications (CNSA 2010)*, Chennai, India, Springer CCIS, edited by Meghanathan et al., Vol. 89, 538-547, Heidelberg, Germany.

171. Wang, Y., G. Attebury, and B. Ramamurthy. 2006. "A Survey of Security Issues in Wireless Sensor Networks." *IEEE Communications Surveys and Tutorials*, 8(2): 2-23.
172. Blom, R. April 1985. "An Optimal Class of Symmetric Key Generation System." In *Proceedings of the EUROCRYPT'84*, 335-338, Paris, France. Springer LNCS, Vol. 209.